\documentclass[trackchanges, twocolumn, twocolappendix]{aastex701}

\usepackage{amsmath,amssymb,amsfonts}
\usepackage{mathrsfs}
\usepackage{booktabs}

\begin{document}

\title{Spectral Decomposition Reveals Surface Processes on Europa}

\author[orcid=0000-0002-1451-6492,gname=Gideon,sname=Yoffe]{Gideon Yoffe}
\affiliation{Department of Earth and Planetary Sciences, Weizmann Institute of Science, Rehovot 76100, Israel}
\email[show]{gidi.yoffe@weizmann.ac.il}  

\author[orcid=0000-0001-9298-8068,gname=Sahar, sname=Shahaf]{Sahar Shahaf} 
\affiliation{Max Planck Institute for Astronomy, K\"{o}nigstuhl 17, D-69117, Heidelberg, Germany}
\email[show]{sashahaf@mpia.de}

\begin{abstract}

Competing processes shape Europa's surface: geological activity replenishes material through resurfacing, while bombardment by charged particles alters surface chemical composition. Each process leaves distinct spectral signatures. We present a novel data-driven analysis of JWST NIRSpec-IFU observations of Europa's leading hemisphere across three observing geometries, targeting nine spectral bands sensitive to water ice, radiolytic products, and volatiles. Through spectral factorization, we isolate the dominant components of spectral variability and reconstruct their spatial distributions. We find that CO$_2$ enrichment extends beyond Tara Regio, and covers multiple chaos units in a lens-like pattern. These CO$_2$-enriched areas co-occur with anomalous ice-texture signatures. Together, these findings suggest that enrichment in volatiles on Europa may reflect retention-favorable near-surface microphysics as well as emplacement, refining how they are interpreted in the context of surface--interior exchange.
This has implications for interpreting the sources and supply rates of extant carbon-bearing species and, ultimately, for assessing Europa's habitability.

\end{abstract}


\section{Introduction} 

Europa, one of Jupiter's Galilean moons, is encased by a geologically young ice shell. This shell probably covers a global, saline subsurface ocean \citep{carr1998evidence}. The habitability of Europa's ocean depends on its chemical composition and redox balance, which is likely shaped by an interplay between internal processes such as water–rock interactions and potential hydrothermal activity, and external inputs of surface-derived oxidants produced by radiation chemistry \citep{chyba2002europa}. While the sub-surface ocean cannot be explored directly, some of its properties can be inferred from those at the surface. Ridges, bands, and widespread chaotic terrains mark its fractured icy shell. These bear evidence of exchange with the interior as well as modification by energetic particle bombardment from Jupiter's magnetosphere \citep{brown2013salts, trumbo2019sodium}. Understanding these processes and quantifying their distinct impact is crucial for determining the ocean's chemical evolution and evaluating Europa's potential to host life.

Radiation from Jupiter's magnetosphere significantly shapes Europa's surface \citep{carlson2009europa}. Since its orbit is located well within this magnetosphere, Europa is continually bombarded by electrons, protons, and heavier ions, which originate mainly from Io's plasma torus \citep{paranicas2001electron, nordheim2022magnetospheric}. These particles impact the surface at energies capable of breaking molecular bonds and driving radiolytic chemistry.
Protons and other ions, particularly oxygen and sulfur, bombard Europa's surface approximately uniformly \citep{nordheim2022magnetospheric}. Electrons, in contrast, bombard the surface in lens-like spatial patterns centered near the equator of each hemisphere, with differing intensities and energies. The trailing hemisphere faces most of the co-rotating plasma flow, dominated by a high flux of lower-energy electrons. The leading hemisphere is exposed to fewer, but more energetic electrons exceeding $\sim$20 MeV \citep{paranicas2001electron}.

The differences in radiation flux produce measurable changes in Europa's surface chemistry. High-resolution near-infrared spectra reveal that the leading and trailing hemispheres exhibit distinct spectral signatures. The trailing hemisphere is dominated by broad absorptions attributed to hydrated sulfuric acid and magnesium sulfate, which are likely formed in situ through radiolytic processing \citep{brown2013salts}.
Radiolytic products such as sulfuric acid (H$_2$SO$_4\cdot n$H$_2$O) and sulfur dioxide (SO$_2$) were also observed in infrared and ultraviolet spectra \citep{carlson1999sulfuric, becker2022mid}.  
In contrast, the leading hemisphere spectra appear smoother, showing water ice features and lacking indicators of hydrated sulfates \citep{ligier2016vlt}. These spectral properties are particularly prominent in geologically young chaos terrain such as Tara and Powys Regiones, which are thought to have experienced recent resurfacing through diapirism or brine extrusion \citep{trumbo2023distribution, villanueva2023endogenous}.

Perhaps the most striking compositional evidence for resurfaced ocean-sourced material on Europa is the detection of irradiated sodium chloride \citep[NaCl;][]{trumbo2019sodium}. Visible and ultraviolet spectra from the \textit{Hubble Space Telescope} (HST) revealed distinct absorptions at 0.23 and 0.45~µm, associated with color centers in irradiated NaCl crystals. These features appear to be spatially linked to the chaos terrains on the leading hemisphere, reinforcing the connection between surface composition and geologically active regions \citep{trumbo2023distribution, villanueva2023endogenous}.

Recent \textit{James Webb Space Telescope} (JWST) observations of the leading hemisphere revealed a carbon dioxide (CO$_2$) absorption doublet near 4.25~µm, spatially concentrated almost entirely within Tara and Powys Regiones \citep{trumbo2023distribution, villanueva2023endogenous}. Its scarcity in older, more heavily irradiated terrains and strong association with geologically young chaos features suggest a subsurface origin. The absence of carbonate or organic precursors in these regions further supports the interpretation that CO$_2$ is either emplaced from the ocean or formed in situ through radiolysis of internally sourced material. 
Surprisingly, a similar spatial pattern is seen in the distribution of hydrogen peroxide \citep[H$_2$O$_2$;][]{trumbo2019h2o2, wu2024europa}. Although one might expect the strongest H$_2$O$_2$ signatures in the cold, ice-rich polar regions, they are actually found at mid-latitudes on the leading hemisphere, concentrated within the warmer chaos terrains. This unexpected distribution suggests that local conditions modulate its longevity.

Still, despite recent observational advances, many aspects of Europa's surface composition remain poorly understood. This is not necessarily due to a lack of high-quality data. Identifying and interpreting variability in spectral measurements is often challenging. Studies usually rely on predefined spectral libraries or band-ratio heuristics, which are inherently limited in detecting unexpected spectral features. This is particularly problematic given that Europa's surface comprises a spatially varying mixture of water ice, salts, radiolytic products, and potentially ocean-derived materials. The local surface conditions shape the relative abundances of these components, and, in the case of resurfaced material, so does the depth and physical environment of the source layer. Hence, a laboratory-produced spectrum cannot fully account for the range of observed chemical constituents under the diverse conditions on Europa's surface. Addressing this challenge requires an inference framework sensitive to subtle spectral structure yet agnostic to the specific origin or form of variability.

\begin{figure*}
\centering
\rotatebox[origin=c]{0}{\includegraphics[scale = 0.75]{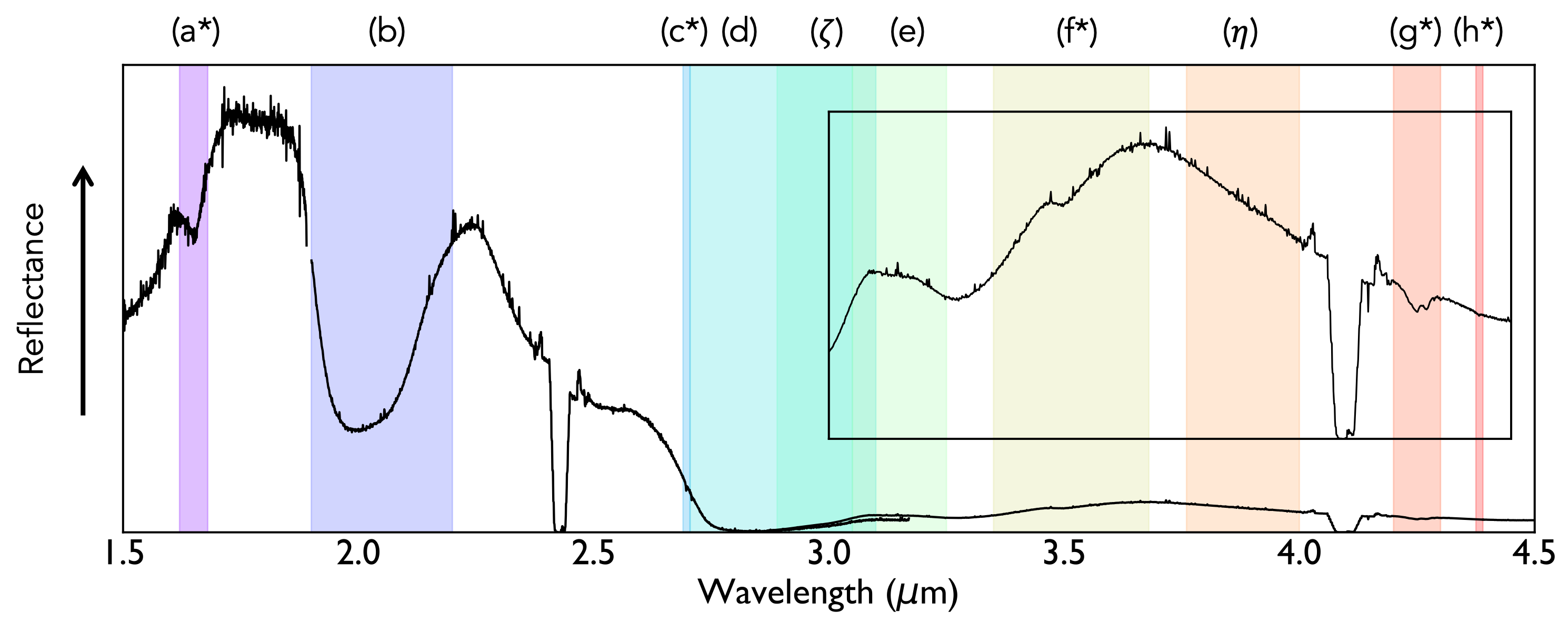}}
  \caption{Selected spectral bands in Europa's normalized NIRSpec-IFU reflectance spectrum. Top: Europa's normalized disk-integrated reflectance spectrum, with seven color-coded spectral bands corresponding to key absorption features. The molecular bands are labeled a--h, and the newly characterized broadband absorption features are labeled with the Greek letters $\zeta$ and $\eta$. Bands marked with an asterisk (a, c, f, g, and h) have been rectified to remove the continuum before further analysis (see Appendix~\ref{sec: obs and reduction}). The positive, narrow, spike-like features are a residual signal due to differences between the Sun and the calibration star.}
     \label{fig: spectrum_integrated}
\end{figure*}

\section{Spectral Factorization of Europa}
\subsection{Data Description}
In this work, we consider the spatially resolved near-infrared observations of Europa from three observing geometries, acquired with the NIRSpec-IFU aboard the JWST. The data were collected using three instrumental configurations: the F290LP filter with the G395H high-resolution grating, which provides continuous coverage from 2.87 to 5.27~µm; the F170LP filter with the G235H grating, which covers 1.66 to 3.17~µm; and the F100LP filter with the G140H grating, which extends coverage down to $\sim$1.0~µm. Together, these modes deliver nearly continuous spectral coverage from $\sim$1 to 5.27~µm at a resolving power of $\sim$2700 and a pixel scale of $\sim$0.1 arcseconds, enabling the identification of surface compositional variability at regional scales. The observations and data reduction details are provided in Appendix~
\ref{sec: obs and reduction}.

Figure~\ref{fig: spectrum_integrated} shows the disk-integrated spectrum of Europa's leading hemisphere, averaged over all spaxels. Nine diagnostic features are evident: (\textbf{a}) the $\sim$1.65~µm crystalline H$_2$O absorption band; (\textbf{b}) the broad $\sim$2~µm H$_2$O absorption band; (\textbf{c}) the narrow $\sim$2.7~µm CO$_2$ combination band \citep[$\nu_1$+$\nu_3$;][]{white2012laboratory}; (\textbf{d}) the $\sim$3~µm segment of the H$_2$O fundamental 3~µm absorption; (\textbf{e}) the 3.1~µm Fresnel surface-reflection peak of H$_2$O ice; (\textbf{f}) the $\sim$3.5~µm H$_2$O$_2$ absorption band;\footnote{The $\sim$3.5~µm band, associated with H$_2$O$_2$, is subjected to significant systematics and, therefore, not analyzed further here. See Appendix~\ref{sec: wiggles} and Figure~\ref{fig: H2O2_trailing_spectrum}.} (\textbf{g}) the 4.25–4.27~µm CO$_2$ $\nu_3$ fundamental band (resolved into a doublet); and (\textbf{h}) a subtler $\nu_3$ absorption near $\sim$4.38~µm attributed to $^{13}$CO$_2$. These bands and continua are associated with active surface processes, including radiolysis \citep{trumbo2019h2o2, wu2024europa} and possible ocean–surface exchange \citep{trumbo2023distribution, villanueva2023endogenous}. We focus on spectral sub-bands centered on these features. 
Lastly, we include two broadband continuum components flanking the $\sim$3.6~µm reflection peak, denoted (\textbf{$\eta$}) and (\textbf{$\zeta$}), which are sensitive to changes in backscattering behavior and peak position.
Several segments (a, c, g, and h) were rectified to suppress continuum variations. See Appendix~
\ref{sec: obs and reduction} for details. Segment (f) is excluded from further analysis due to systematic contamination (see Figure~\ref{fig: H2O2_trailing_spectrum} and appendices below). 

The selected bands, shown in the bottom panels of the figure, are arranged in a matrix, with each row corresponding to a spaxel. See a quantitative definition in equation~\ref{eq: flattening}.  This matrix (denoted $S$ in the following) represents a specific measurement epoch, where an IFU data-cube of Europa was obtained at some given observing angle. The decomposition of this matrix provides the foundation for the analysis described below.

\subsection{Spectral-Spatial Decomposition}

To overcome the challenges described above, we developed a data-driven framework based on spatially resolved near-infrared spectra of Europa's leading hemisphere. These spectra were acquired with the NIRSpec integral field unit (IFU) aboard JWST \citep{boker2022near}. The IFU simultaneously captures spectra from multiple locations across the surface, each sampling slightly different local conditions and producing slightly different spectral signatures.

Our analysis assumes that the reflected spectrum varies across the surface as a function of a few physical and geometric parameters. While individual spectra may exhibit complex structures, their variability can be captured if the spectrum responds smoothly to local parameter changes. Therefore, each spectrum can be approximated as a linear combination of a limited set of basis functions. A recent study has developed this factorization to analyze stellar spectra in the context of precision radial velocity measurements \citep{Shahaf2023, Shahaf2025}. Following their definitions, we refer to these functions as `\textit{principal spectra}.'
Under this assumption, the spectrum observed at each IFU pointing can be written as 
\begin{equation} 
S_{k\lambda} = u^{(0)}_{k} v^{(0)}_\lambda + u^{(1)}_{k} v^{(1)}_\lambda  + u^{(2)}_{k} v^{(2)}_\lambda + \dots, 
\label{eq: factorization}
\end{equation} 
where $v^{(i)}$ are the principal spectra, and $u^{(i)}$ their corresponding spatial scaling coefficients, which we refer to as `\textit{spatial modes}.'  The $\lambda$ index denotes the wavelength-axis index, and $k$ represents the index of the IFU spaxel (spatial pixel). The latter is a one-dimensional running index mapped to each spaxel's two-dimensional position. See Data Description and Preprocessing.

While neither the principal spectra nor their scaling coefficients are known a priori, the bilinear structure of the proposed decomposition is well-suited for singular value decomposition \citep[SVD;][]{golub2013matrix, Shahaf2025}, enabling their derivation. Therefore, we treat $S_{k \lambda}$ as a matrix, decomposed via
\begin{equation}
    S = U \Sigma V^\top.
    \label{eq: SVD}
\end{equation}
This process yields a set of left and right singular vectors stored in the orthonormal matrices $U$ and $V$, respectively. The columns of these matrices correspond, up to a multiplicative factor, to the spatial modes and principal spectra from equation~(\ref{eq: factorization}). The columns are sorted by the proportion of the signal each describes, which is stored in the diagonal singular-value matrix, $\Sigma$.

Each $V$ column defines a distinct, linearly independent principal spectrum, representing a characteristic variation in spectral shape across the dataset. The zeroth-order principal spectrum captures the dominant, average spectral profile, whereas higher-order modes represent deviations from it. These deviations reflect surface constituents, geological features, exogenic processing, and potential instrumental effects.
As an orthogonal variance decomposition, the modes need not map uniquely onto distinct physical processes.
The corresponding $U$ columns represent the spatial modes, describing how the contribution of each principal spectrum varies across Europa's leading hemisphere. Each spatial mode is initially derived as a one-dimensional vector over the IFU spatial index and is then reshaped into a two-dimensional map of spaxel positions on the detector, as described in the Appendices.

If the target is observed from multiple viewing geometries, the spectra from all observations can be decomposed jointly by stacking the corresponding data matrices,
\begin{equation}
    \begin{bmatrix}
    S_1 \\
    \vdots \\
    S_n
    \end{bmatrix}
    = 
    \begin{bmatrix}
    U_1 \\
    \vdots \\
    U_n
    \end{bmatrix}
    \Sigma \, V^\top.
    \label{eq: SVD multi hemisphere}
\end{equation}
where $S_i$ is the spaxel–by–wavelength matrix for the $i$\textsuperscript{th} observing geometry. In this formulation, the right singular vectors $V$ define a set of principal spectra that are common to all observations, while the left singular vectors $U_i$ encode the geometry-dependent spatial weighting of those spectra on the detector.

This construction separates intrinsic spectral structure from viewing-dependent effects. Projection, illumination, and instrumental differences are absorbed into the spatial coefficients, while the spectral basis is constrained jointly by all geometries. As additional hemispheres, epochs, or instruments are incorporated, degeneracies inherent to any single projected view are progressively reduced, enabling a surface-consistent spectral representation. It also enables direct cross-geometry comparison of variability, while boosting the effective signal-to-noise ratio through pooled spaxel constraints.

\begin{figure*}
\centering
\rotatebox[origin=c]{0}{\includegraphics[scale = 0.585]{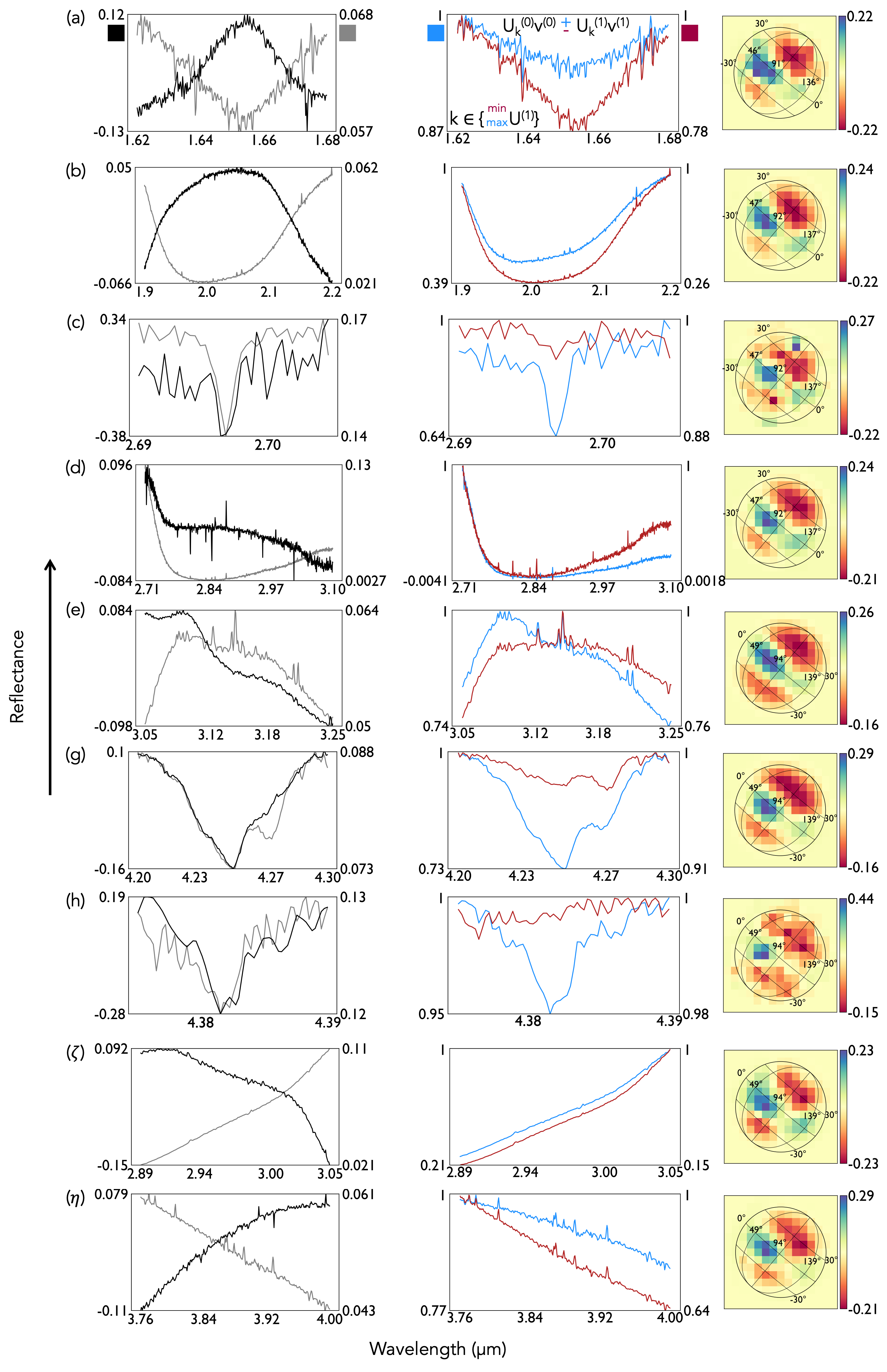}}
  \caption{Spectral and spatial modes of Europa's diagnostic bands. Left: zeroth and first spectral modes, $v^{(0)}$ (gray) and $v^{(1)}$ (black) (see Section~\ref{sec: principal spectra}). The left and right ordinates give the true extrema of $v^{(0)}$ and $v^{(1)}$, marked by gray and black squares in panel (a), while the plotted curves are rescaled for clarity. Middle: spectra from the spaxels where the first spatial mode ($u^{(1)}$) peaks (blue) and reaches its minimum (red), each scaled so its own maximum is 1 (minima shown relative to that peak). The left ordinate corresponds to the blue spectrum and the right ordinate to the red spectrum, as indicated by the colored squares in panel (a); both spectra are rescaled for presentation. Right: first spatial mode, $u^{(1)}$, corresponding to the principal spectra shown in the middle and left panels (see Section~\ref{sec: spatial modes}).}
     \label{fig: spectrum_decomposition}
\end{figure*}

Figure~\ref{fig: spectrum_decomposition} shows the first spectral and spatial modes ($v^{(1)}$ and $u^{(1)}$, respectively) derived from the decomposition of each of the selected bands for the face-on observation of the leading hemisphere, which are fully covered by the NIRSpec spectral range. For this purpose, we used the special single-viewing-angle case shown in equation~(\ref{eq: SVD}). In the following, we present a multi-hemisphere decomposition, using equation~(\ref{eq: SVD multi hemisphere}), for spectral bands for which several viewing angles of Europa are available. 

\section{Trends in Europa's Reflectance}
\label{sec: principal spectra}
First, we interpret the principal spectra extracted from each band to identify the dominant sources of variability. The results, shown in Figure~\ref{fig: spectrum_decomposition} and discussed below, yield interpretable signatures that can be compared to laboratory diagnostics.

\subsection{Europa's principal spectra}
We restrict our analysis to the first two principal spectra, $v^{(0)}$ and $v^{(1)}$. Higher-order modes are largely dominated by instrumental sub-sampling effects and are removed during the pre-processing stage (see Appendix~\ref{sec: obs and reduction} and figures therein). The retained components capture the leading-order spectral variability. The zeroth-order principal spectrum represents the global mean spectrum and encodes the baseline surface composition, while the first-order captures the dominant patterns of spectral variability across the surface.

The singular values and singular vectors obtained from the decomposition are subject to uncertainty because the analysis is performed on spectra contaminated by measurement noise. Because singular vectors are normalized, this uncertainty manifests as a rotation within the corresponding vector subspaces rather than a change in their norms. In practice, this implies that the orientations of the extracted spectral and spatial modes are perturbed relative to their noise-free counterparts. We explicitly derive and quantify these uncertainties by propagating the per-spaxel measurement errors through the SVD, yielding bounds on the angular deviations of the leading modes and corresponding uncertainties in their spatial structure. The formalism and resulting bounds are developed in Appendix~\ref{app: error propagation} (see also \citealt{Shahaf2025}). The stability of the principal spectra in each band is quantified in Table~\ref{tab:svd_noise} and discussed per band below.

We sample a large fraction of the near-infrared spectrum across all available datasets, with primary sensitivity to variability associated with H$_2$O ice and CO$_2$-bearing phases. Accordingly, the dominant modes extracted by the decomposition predominantly trace variations in these components. The observed properties of each band are discussed below.

\subsection{Interpreting the derived terms}
The principal spectra isolate the dominant variability across Europa's near-infrared reflectance, but such variations do not uniquely map onto a single microphysical parameter. Accordingly, the spectral trends discussed below should not be interpreted as direct measurements of specific properties. Instead, they constrain the effective scattering and optical path regimes of near-surface ice. While individual bands are degenerate, their joint behavior across multiple diagnostics substantially restricts admissible physical states. We therefore interpret the first-order spectral modes as tracing ice texture, without assigning a unique microphysical cause.

{\bf H$_2$O ice:} The 1.65~µm overtone band of the O--H stretching mode is a diagnostic of crystalline water ice down to depths of order hundreds of microns, manifesting in ordered ice at low temperatures \citep{grundy1998temperature, mastrapa2008optical}. It is strongly reduced in amorphous-dominated spectra and weakens with increasing temperature, becoming difficult to detect by $\sim$120--130~K \citep{grundy1998temperature, mastrapa2008optical, hansen2004amorphous}.
The left panel of Figure~\ref{fig: spectrum_decomposition}a shows $v^{(0)}$ and $v^{(1)}$, which is morphologically similar but opposite in sign. Adding it weakens the mean absorption feature ($v^{(0)}$), while subtracting it deepens it.

The $\sim$2~µm H$_2$O absorption band is the strongest non-saturated near-infrared water-ice band in our data and is widely used to trace ice abundance and effective grain size on icy satellites \citep[e.g.,][]{grundy1998temperature, filacchione2012saturn, clark2012surface}. Figure~\ref{fig: spectrum_decomposition}b shows that $v^{(1)}$, similarly to that of the 1.65~µm feature, primarily controls the depth of the absorption. It also introduces a slight asymmetry in its profile, which is consistent with the temperature-dependent shape changes of this band observed in laboratory spectra \citep{grundy1998temperature, mastrapa2008optical}.

The 2.71–3.10~µm segment samples the short-wavelength wing of the fundamental 3~µm H$_2$O absorption. Because the fundamental is extremely strong, this region is dominated by very shallow layers and is particularly sensitive to surface microtexture and phase \citep{mastrapa2009optical}. In Figure~\ref{fig: spectrum_decomposition}d, $v^{(1)}$ steepens the 3~µm wing and shifts the minimum closer to the nominal H$_2$O fundamental.

Figure~\ref{fig: spectrum_decomposition}e presents the 3.1~µm Fresnel peak. This feature arises from surface reflection at the fundamental O--H stretching band of water ice and is sensitive to the ice phase of the uppermost layer \citep{hansen2004amorphous}. Amorphous ice produces a single broad peak centered near 3.1~µm, while crystalline ice exhibits a narrower Fresnel component whose band center is shifted and can be accompanied by a secondary shoulder near 3.2~µm \citep{hansen2004amorphous, mastrapa2009optical, stephan2021vis}. The detailed double-peaked morphology is not a simple diagnostic of crystallinity, as its 3.2~µm shoulder strength depends on microphysical properties and temperature \citep{hansen2004amorphous, mastrapa2009optical, stephan2021vis}, but remains reliably linked to the surface ice phase.
The middle panel of Figure~\ref{fig: spectrum_decomposition}e shows how $v^{(1)}$ isolates this component. When added to or subtracted from the mean spectrum, the resulting profiles bracket laboratory spectra of amorphous and crystalline surface ice (blue and red curves, respectively) in the uppermost few microns \citep{hansen2004amorphous, mastrapa2009optical}. 

Figure~\ref{fig: spectrum_decomposition}$\zeta$ and $\eta$ presents broadband modulations between $2.9–3.05$~µm and 3.76–4.0~µm. These appear as smooth drops in reflectance bracketing the broad 3.6~µm peak, shaped by strong 3~µm H$_2$O absorption and multiple scattering in the ice \citep{hansen2004amorphous, stephan2021vis}. Their corresponding $v^{(1)}$ spectra primarily modulate the width and slope of the `shoulders' falling blue- and red-wards from the 3.6~µm reflectance peak (see the inset in Figure~\ref{fig: spectrum_integrated}).

{\bf CO$_2$:} 
We examine the three CO$_2$–related bands highlighted above: the 2.7~µm $\nu_1+\nu_3$ combination band, the $\sim$4.2–4.3~µm CO$_2$ $\nu_3$ fundamental, and the 4.38~µm $^{13}$CO$_2$ $\nu_3$ band. The 2.7~µm feature (Figure~\ref{fig: spectrum_decomposition}c) provides an additional tracer of solid CO$_2$ and is consistent with CO$_2$ molecules embedded in H$_2$O–dominated and mixed matrices \citep[e.g.,][]{white2012laboratory}. In the $\nu_3$ region (Figure~\ref{fig: spectrum_decomposition}g), the broad 4.25~µm component dominates the disk-integrated spectrum, as expected for CO$_2$ in more weakly bound or mixed environments such as association with non-ice materials \citep{white2012laboratory, villanueva2023endogenous, trumbo2023distribution, schiltz2024characterization}. The first singular spectrum, $v^{(1)}$, is approximately a version of this profile but lacks the narrow 4.27~µm peak. Adding $v^{(1)}$ to the mean spectrum therefore enhances the 4.27~µm component while suppressing the broader 4.25~µm absorption, and vice versa. 

The $^{13}$CO$_2$ band appears as a narrow absorption centered at 4.38~µm and has recently been detected on Europa’s surface \citep{cartwright2025jwst}. The corresponding singular spectrum and spatial mode are shown in Figure~\ref{fig: spectrum_decomposition}h; as discussed in the Error Estimation section, the detection significance is limited, and we use this band only to confirm that the $^{13}$CO$_2$ distribution follows the total CO$_2$ within the noise limits.

\section{Spatial Trends} 
\label{sec: spatial modes}
We next examine the associated spatial modes directly at the IFU spaxel level. The right panels of Figure~\ref{fig: spectrum_decomposition} show the first-order spatial modes, $u^{(1)}$, derived from the detector-plane data for each spectral band. These maps reveal that the dominant spectral variability is spatially localized, with distinct bands exhibiting structures that differ in extent and morphology. In several cases, enhanced or suppressed spectral signatures are confined to specific regions rather than distributed uniformly across the surface, indicating that spectral changes are geographically structured and band-dependent.

The characteristic spatial scales and locations of these features vary between diagnostics, suggesting that different physical or microphysical processes contribute to the observed variability across bands. These spaxel-level maps therefore provide a direct view of the intrinsic spatial structure present in the data. In the following sections, we quantify and compare these spatial patterns for different spectral bands and viewing geometries.

\subsection{Approximation to spherical harmonics}

The finite spatial resolution of the IFU allows the derived spatial modes to be represented compactly. Rather than working directly in detector-pixel space, we parameterize each spatial mode $u^{(k)}$ using a truncated expansion in real spherical harmonics defined on the surface of Europa and orthographically projected onto the detector plane. Specifically,
\begin{equation}
u^{(1)} \simeq \sum_{l,m} a_{lm}\, R_{lm}(\hat{\mathbf{n}}_k),
\end{equation}
where $k$ indexes the IFU spaxels as described in equation~(\ref{eq: flattening}), and $\hat{\mathbf{n}}_k$ denotes the detector-projected coordinates on spaxel $k$ under the adopted viewing geometry. The real spherical harmonics $R_{lm}$ are evaluated on the surface coordinates, while the viewing direction enters through the mapping between spaxels and surface normals.
The approximation holds in the weighted least-squares sense, and the coefficients $a_{lm}$ are obtained via linear regression. A detailed description of the projection and fitting procedure is given in Appendix~\ref{app: model_Ylm}.
Spherical harmonics provide a convenient framework for modeling spatial modes under multiple viewing geometries. By fitting a common set of harmonic coefficients to observations obtained under different projections and point-spread functions, the same underlying surface structure can be constrained simultaneously. When restricted to a single projected hemisphere, the harmonic components are not uniquely identifiable, as different modes can produce similar patterns in the image plane. In this regime, the representation is intentionally non-unique and is not interpreted as a unique surface reconstruction. Nevertheless, the truncated expansion remains useful: it enforces a smooth, scale-ordered, and sparse description of spatial variability on IFU-resolved scales and provides a parameterization that can be consistently extended as additional viewing geometries or instrumental responses are incorporated.
\begin{figure*}[t!]
\centering
\rotatebox[origin=c]{0}{\includegraphics[scale = 0.25]{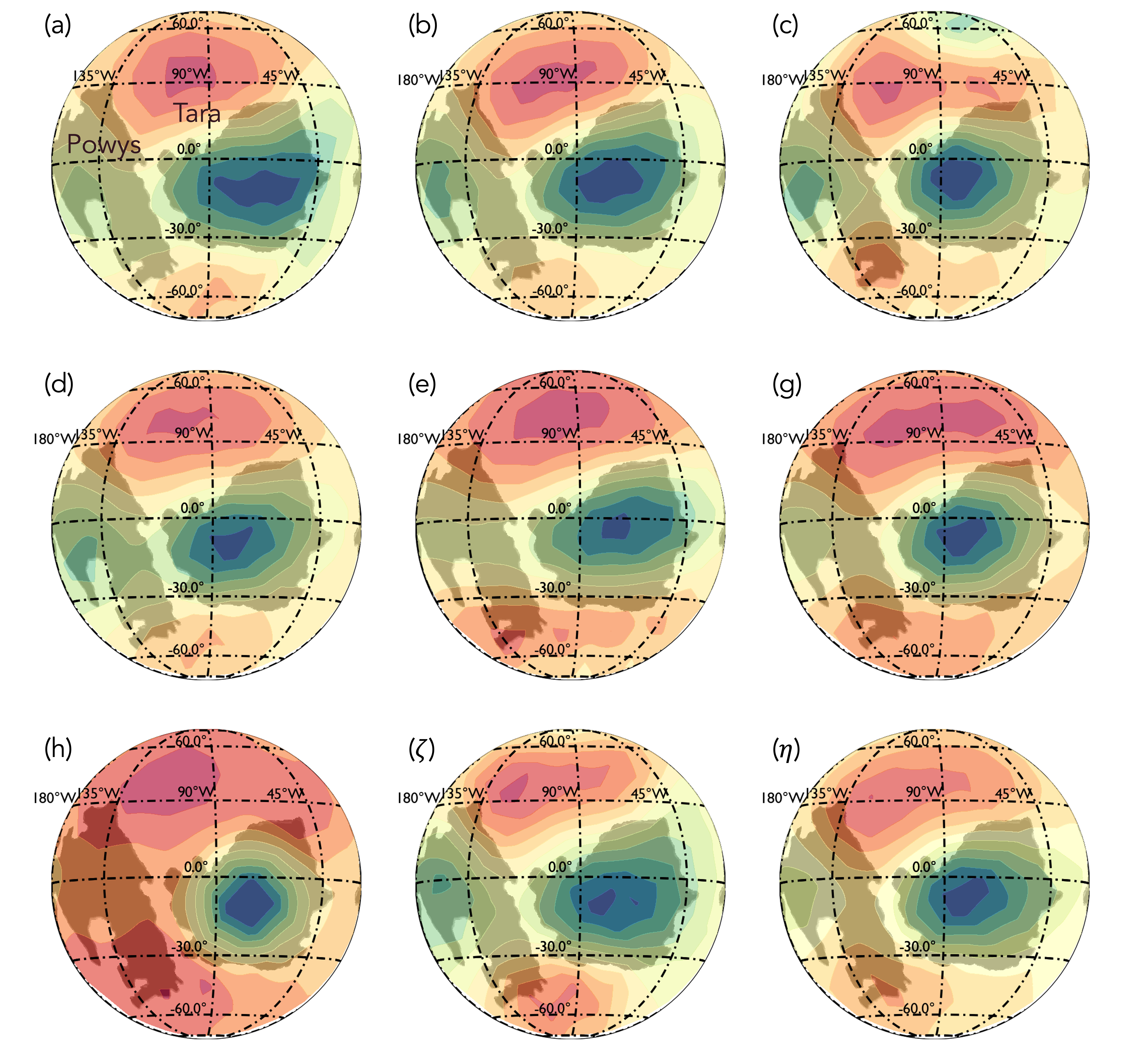}}
  \caption{Qualitative projections of best-fit spherical harmonic models for the first-order spatial modes, $u^{(1)}$, associated with four molecular bands (labeled a-h, excluding the H$_2$O$_2$ feature) and two newly characterized broadband continuum band-widening features (labeled $\zeta$,$\eta$), overplotted on a binary mask of Tara (black; right-hand feature) and Powys (black; left-hand feature) Regiones. Colors indicate the signed values of $u^{(1)}$ (blue: positive; red: negative), scaled to the extrema of each fitted mode; corresponding numerical ranges are shown in Figure~\ref{fig: spectrum_decomposition}.}
     \label{fig: projections}
\end{figure*}
\subsection{Spatial modes of the leading hemisphere}

\begin{figure} [t!]
\centering
\rotatebox[origin=c]{0}{\includegraphics[scale = 0.44]{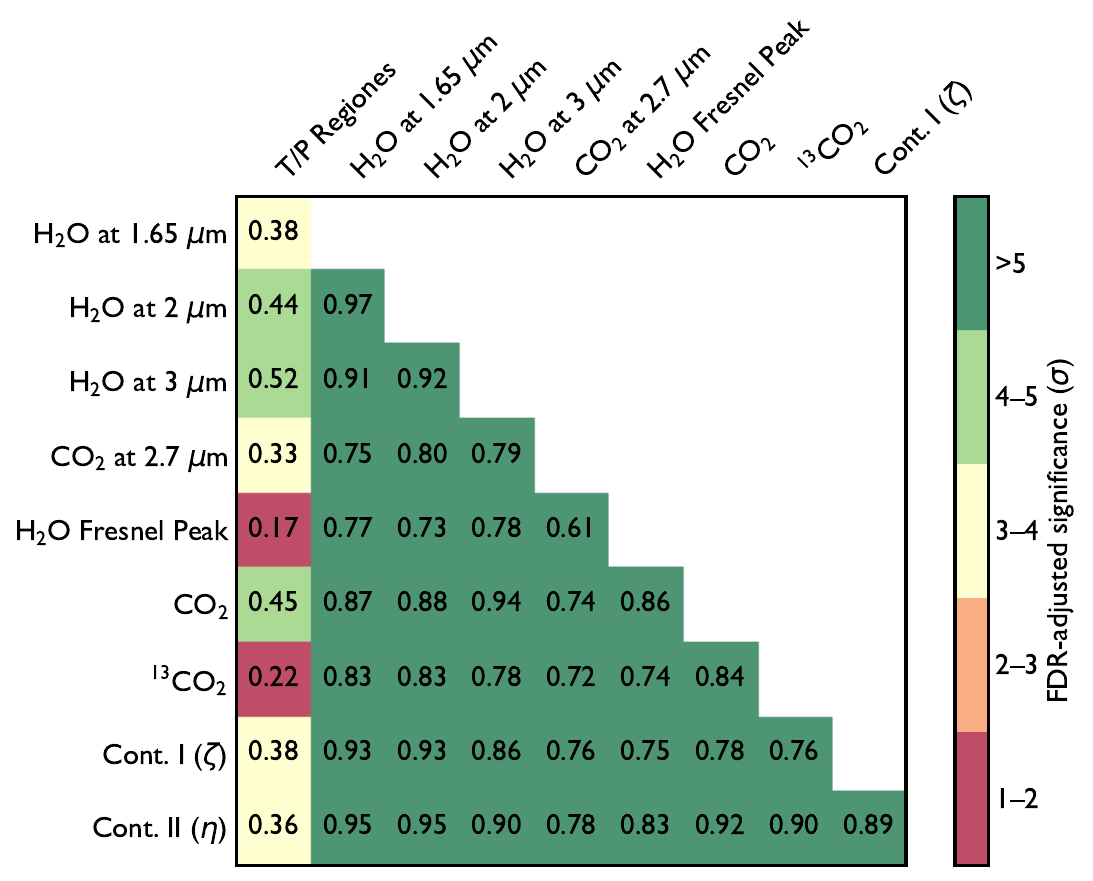}}
  \caption{Significance matrices of Pearson correlations between best-fit spherical-harmonic coefficients of the first-order spatial modes across spectral bands and rescaled and orthographically-projected terrain masks leading hemisphere with a binary mask of Tara and Powys Regiones (see the Appendices). Lower-triangular panels report pairwise Pearson correlation coefficients (text) between bands; color encodes statistical significance after false discovery rate correction \citep{benjamini1995controlling}. The color bar shows the corresponding Z-score derived from FDR-adjusted $p$-values; greener indicates higher significance.
\textbf{Note}: Correlation values displayed here are accurate up to a sign, consistent with the inherent sign indeterminacy of SVD \citep{golub2013matrix}.}
\label{fig: corr_mat}
\end{figure}

Figure~\ref{fig: projections} shows the leading-hemisphere projection of the fitted model, overlaid on a binary map of Tara and Powys Regiones; the fits reproduce the observed modes with $R^2>0.95$ in all bands. Figure~\ref{fig: corr_mat} summarizes pairwise correlations between the best-fit spherical-harmonic coefficient vectors across bands. The correlations are statistically robust: the water-ice bands co-vary tightly, several also co-vary with the CO$_2$ bands, and multiple bands correlate with a binary mask of Tara and Powys Regiones, consistent with modes tied to the dominant chaos terrains.

Most spatial modes associated with water-ice are concentrated in southern latitudes, with the strongest expression in Tara Regio and a secondary enhancement in Powys Regio. The 3.1~µm Fresnel peak, associated with crystalline ice in the uppermost surface, extends furthest north, reaching to ${\sim}15^\circ$N and trending southwestward to ${\sim}30^\circ$S  (Figure~\ref{fig: projections}e). In contrast, the weakening of the 1.65~µm and $\sim$2~µm absorption bands (Figure~\ref{fig: projections}a and b) is concentrated in southern Tara Regio, with a weaker counterpart in Powys Regio, and is accompanied by reduced reflectance on the 3~µm wing (Figure~\ref{fig: projections}d).

Both shoulders of the 3.6~µm H$_2$O reflectance peak (Figure~\ref{fig: projections}$\eta$ and $\zeta$) show a simultaneous increase in backscattering over Tara Regio, with a weaker expression over Powys Regio, broadly tracking the behavior of the other water-ice bands. Taken together, the enhanced backscattering and the shallowing of shorter-wavelength ice absorptions over the dominant chaos units are consistent with smaller or rougher grains, increased porosity, or a combination of both \citep[e.g.,][]{shkuratov2001opposition, hapke2002bidirectional, hapke2008bidirectional, muinonen2010coherent}.

The observed spatial pattern supports a previously observed clear latitudinal asymmetry in ice texture, with the southern Tara Regio standing out as an anomalous region that hosts enhanced crystallinity, more heterogeneous ice, and textural characteristics distinct from those of its surroundings \citep[e.g.,][]{ligier2016vlt, cartwright2025jwst}. Outside the chaos terrains, the phase of surface and near-surface ice is dominated by different processes, with the former predominantly amorphous and the latter predominantly crystalline.

The increase in the 4.25~µm CO$_2$ absorption band (Figure~\ref{fig: projections}g), together with the enhancement of the 2.7~µm CO$_2$ $\nu_1+\nu_3$ combination band (Figure~\ref{fig: projections}c), is more sharply localized, centered in southwestern Tara Regio with weakened expressions in Powys Regio. $^{13}$CO$_2$ is confined to an even narrower zone within the same area (Figure~\ref{fig: projections}h). However, the significance of the latter is limited, as its first-order principal mode is moderately susceptible to noise-induced perturbations, and we consider it tentative (see Appendix~\ref{app: error propagation}). The weaker expression of variability patterns in Powys Regio may partly reflect its proximity to the limb during observation, where foreshortening and line-of-sight mixing blend larger surface areas into each spaxel, whereas Tara Regio lies closer to the sub-observer point; this ambiguity is mitigated when multiple geometries are analyzed jointly, as shown below.

\subsection{Bands observed from multiple angles} \label{sec: multi_geometry_fit}
\begin{figure*}[t!]
\centering
\rotatebox[origin=c]{0}{\includegraphics[scale = 0.24]{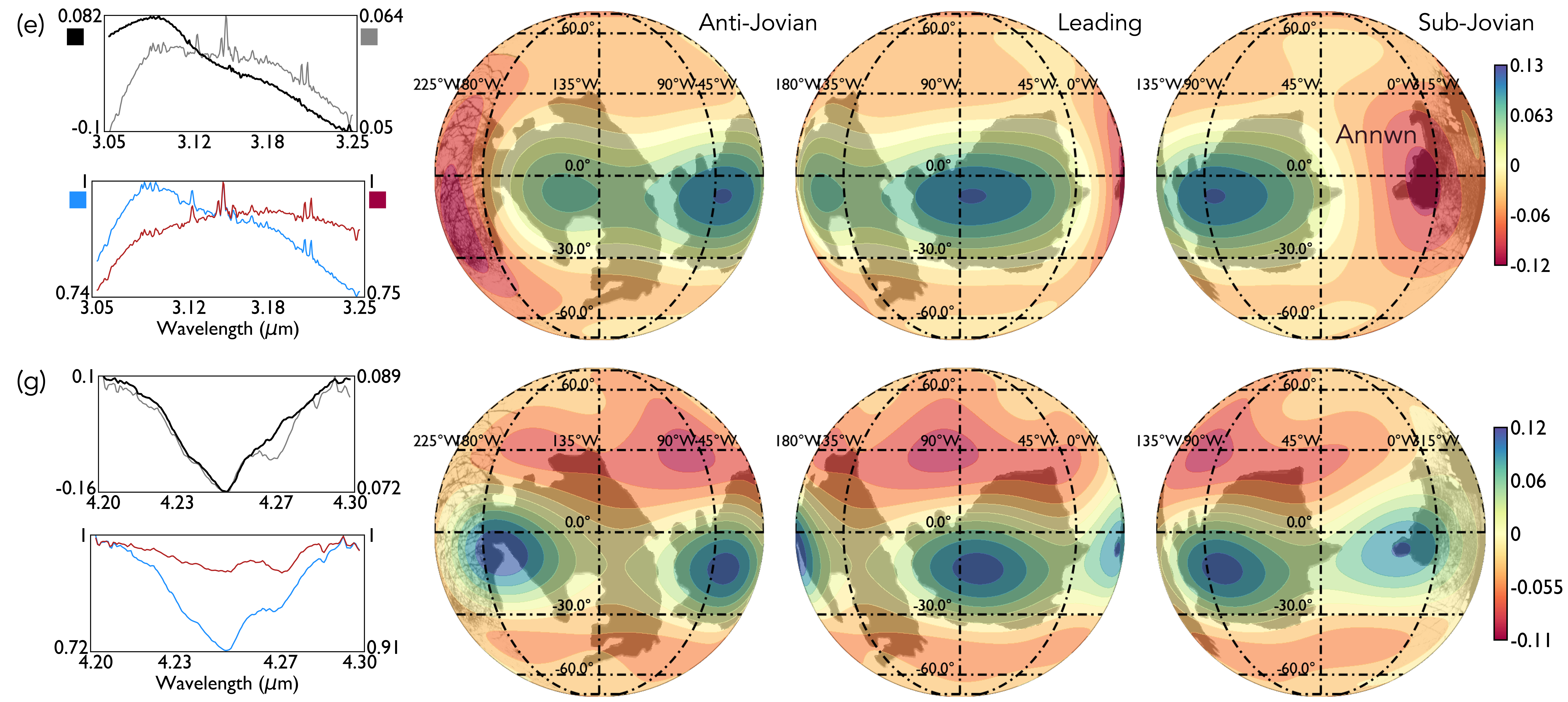}}
  \caption{Qualitative projections of the best-fit three-dimensional spherical-harmonic model to the jointly decomposed first spatial mode, $u^{(1)}$, derived from three viewing geometries of Europa's leading hemisphere. (e) Water-ice 3.1~µm Fresnel peak ($\ell_{\rm{max}}=6$). (g) CO$_2$ absorption doublet near 4.25~µm ($\ell_{\rm{max}}=7$). Shaded outlines mark Tara, Powys, and the leading-hemisphere part of Annwn Regiones. The left panels show jointly decomposed spectral modes ($v^{(0)}$ and $v^{(1)}$ in gray and black, respectively) and spectra from the spaxels where the joint spatial mode peaks (blue) and reaches its minimum (red).
}
  \label{fig: fresnel_co2_mult_proj}
\end{figure*}

In this work, we consider two bands observed from multiple viewing angles: the 3.1~µm water-ice Fresnel peak (Figure~\ref{fig: fresnel_co2_mult_proj}a) and the CO$_2$ absorption doublet near 4.25~µm (Figure~\ref{fig: fresnel_co2_mult_proj}b). Both were repeatedly observed on the leading hemisphere in the F290LP/G395H filter-grating configuration. For these bands, we obtain a single set of principal spectra common to all viewing geometries, as given by equation~(\ref{eq: SVD multi hemisphere}), allowing the corresponding spatial modes to differ between projections. 
The left panels of Figure~\ref{fig: fresnel_co2_mult_proj} present the jointly-derived principal spectra, together with the spherical-harmonic approximation of their corresponding spatial modes. This parameterization enables direct comparison of spatial structure between viewing geometries.
The procedure for analyzing multiple viewing angles is described in Appendix~\ref{app: model_Ylm} (see also Figure~\ref{fig: multigeometry_fit_example}, which presents the explicit spatial modes).

The joint multi-geometry analysis shows enhanced 4.25~µm CO$_2$ absorption appears as localized features that recur in distinct chaos terrains, including regions near the anti-Jovian edge of Powys Regio and the leading-hemisphere edge of Annwn Regio, just beyond the sub-Jovian meridian. In contrast, the 3.1~µm Fresnel peak exhibits a broader, double-lobed morphology, correlated with both Tara and Powys Regiones, with its maximum occurring at Tara Regio. Notably, the Fresnel signal does not peak near the sub- or anti-Jovian points, where the surface albedo decreases \citep{mergny2025blinking}.

For completeness, we also jointly analyze three bands in the $\sim$2--3~µm range in the two geometries where they are observed, the anti-Jovian and leading hemispheres; the results are presented in Appendix~\ref{app_multi}. These bands exhibit different spatial distributions within the geometries analyzed here. In particular, their detailed morphologies around Powys Regio are not identical, although these differences are tentative, given the proximity to the limb and incomplete sampling toward the trailing hemisphere. The 2.7~µm CO$_2$ band is confined to a discrete lens near the anti-Jovian meridian of Powys Regio and is co-located with the 4.25~µm CO$_2$ absorption. The 3~µm H$_2$O fundamental shows similar spatial structure to the 4.25~µm CO$_2$ absorption, whereas the $\sim$2~µm band shallowing follows terrain darkening and appears to peak beyond the anti-Jovian meridian. While such variations may reflect distinct surface microphysics, a comprehensive physical interpretation requires broader spatial coverage, including the trailing hemisphere, and is deferred to future work.

Three additional bands in the $\sim$3--5~µm range have multi-geometry coverage from three viewing angles: The $^{13}$CO$_2$ band is only marginally detected in the near face-on observation and is noise-dominated in the remaining views, and was not considered for further analysis. The other two are the shoulders of the 3.6~µm water ice reflectance peak, for which some leading-hemisphere viewing angles are affected by backscattering from darkened terrain adjacent to the trailing hemisphere. These signatures likely reflect radiolytic processing into hydrated sulfates \citep{carlson1999sulfuric, carlson2005distribution}. Since the driving physical mechanism may differ from that producing a morphologically similar signature within Tara Regio and may require additional observational constraints, we also defer their joint analysis to future work and consider only a single viewing angle.

\section{Discussion} \label{discussion}

This study reveals hemisphere-scale contrasts in Europa's near-infrared spectrum. On the leading hemisphere, spectral variability distinguishes chaos terrains from their surroundings. These differences are expressed through variations in water-ice bands and volatile absorptions, which together highlight systematic contrasts in ice texture and composition.

\subsection{Water ice on Europa's leading hemisphere}
The above-mentioned water-ice diagnostics show differences between the chaos terrains, most prominently southern Tara Regio, and their surrounding area. Tara Regio exhibits a pronounced double-lobed 3.1~µm Fresnel morphology (band~e), relatively shallow 1.65~µm and 2~µm H$_2$O bands (bands~a and b, respectively), a deepened, reddened long-wavelength wing of the 3~µm H$_2$O complex (band~d), and broadened 3.6~µm shoulders (bands $\eta$ and $\zeta$). 
This is consistent with a more crystalline near-surface veneer in the uppermost microns, given laboratory spectra in which crystalline H$_2$O displays a pronounced double-lobed Fresnel feature under comparable conditions \citep{mastrapa2009optical,hansen2004amorphous,stephan2021vis}. 
This interpretation is consistent with previous Europa work that uses the 3.1~µm Fresnel peak as a crystallinity indicator in Tara and Powys Regiones \citep{ligier2016vlt, cartwright2025jwst}.

In contrast, the joint weakening of the 1.65~µm and 2~µm bands, which probe deeper volumetric ice, is consistent with radiative-transfer regimes in which the effective optical path length of H$_2$O is reduced. Increased porosity or roughness, reduced effective grain size, and mixing with non-ice material can enhance multiple scattering and diminish apparent band depths at a fixed composition \citep{shkuratov1999model, hapke2002bidirectional, hapke2008bidirectional, filacchione2012saturn, stephan2021vis}. Laboratory spectra show that the 1.65~µm band weakens with increasing temperature and with decreasing crystalline fraction, whereas the 2~µm band depth is primarily controlled by particle size and varies comparatively weakly with temperature over under Europa-like conditions \citep{grundy1998temperature, mastrapa2008optical, stephan2021vis}. The observed asymmetry of the 2~µm profile is also consistent with an increase in temperature, as reported in laboratory spectra \citep{grundy1998temperature, mastrapa2008optical}.

The two broadband shoulders of the 3.6~µm H$_2$O reflection peak are not uniquely diagnostic. Similar broadband continua are influenced by ice temperature and abundance, effective grain size, and mixing with darker materials on icy satellites \citep{hansen2004amorphous, filacchione2012saturn, clark2012surface}.
The spectral-spatial decomposition results in simultaneous broadening of both shoulders in the chaos terrains (see Figure~\ref{fig: spectrum_decomposition}).
Such an increase is expected in interface-rich, highly scattering media in which coherent-backscatter contributions can become more pronounced and can modify continuum slopes even if bulk reflectance decreases \citep{hansen2004amorphous, stephan2021vis, hapke1990coherent, hapke2002bidirectional, muinonen2010coherent, shkuratov2001opposition}.

Considering all bands together, these spectral variations delineate an anomalous ice texture localized to chaos units. Their spatial distribution is not naturally expected from irradiation or diurnal heating alone, which to first order vary smoothly with latitude and illumination geometry, nor from slow compaction acting over much longer length and time scales \citep{mergny2024gravity}.

\subsection{Expected evolution of near-surface water-ice}
\begin{figure*}
\centering
\rotatebox[origin=c]{0}{\includegraphics[scale = 0.42]{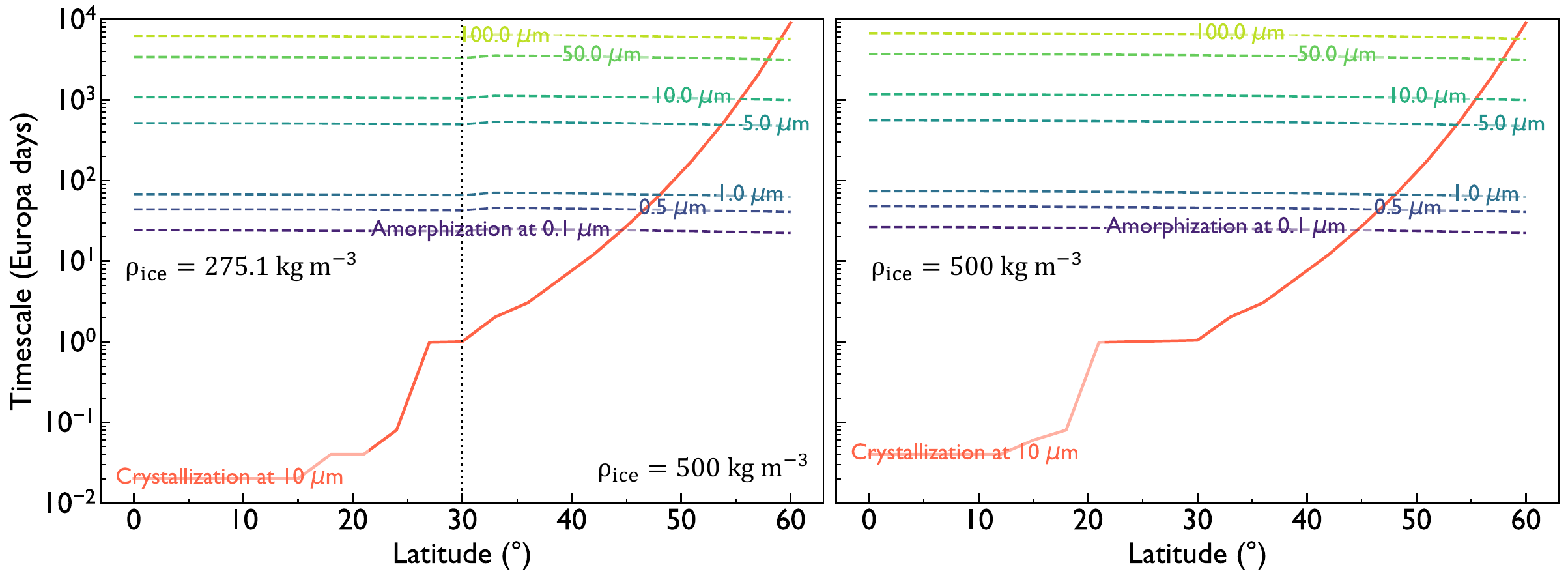}}
  \caption{Latitude-dependent timescales for ice amorphization and crystallization of Europa's leading hemisphere's near-surface ice. Timescales are computed independently for each process and are shown for (left) porous ice at latitudes $<30^\circ$ and otherwise compact, and (right) compact ice. Amorphization rates (dashed lines) are plotted for depths of 0.1--100~µm (color-coded by depth), while a representative crystallization rate is shown at 10~µm depth (solid red line).}
     \label{fig: amorphization_crystallization_rates}
\end{figure*}

We modeled the thermal and radiolytic evolution of near-surface ice on Europa to assess whether ambient surface conditions can explain the crystallinity gradient.
By coupling depth- and time-resolved temperature profiles \citep{ashkenazy2019surface} with radiation-driven amorphization rates, we quantified the competition between recrystallization and radiolytic amorphization under realistic surface conditions \citep{mitchell2017porosity, yoffe2025fluorescent}. A full description of the process, assumptions, and results is provided in Appendix~\ref{app_ice}. The results are illustrated in Figure~\ref{fig: amorphization_crystallization_rates}.

The model shows that ice crystallization across the lower-albedo chaos outpaces amorphization at latitudes below 30$^\circ$. This estimate was obtained assuming two scenarios: (1) a global porosity value of ${\sim}\,0.5$, obtained by empirical estimates of Europa's near-surface regolith mean thermal inertia \citep{thelen2024subsurface, mergny2025blinking}, and (2) distinct low- and high-latitude porosity values of $\sim0.5$ and $\sim0.7$, respectively, as inferred from reduced thermal inertia values at Tara Regio (as observed by ALMA; see \citealt{thelen2024subsurface}). We defined low latitudes as the range 0$^\circ$--30$^\circ$. In both cases, low-latitude ice remains crystalline under ambient surface conditions. Increased porosity and reduced albedo may further accelerate crystallization, leading to the observed localized elevated crystallinity patterns at the chaos units.

The difference between the two scenarios presented in Figure~\ref{fig: amorphization_crystallization_rates} is the expected crystallization timescale. The modeling suggests that this timescale shortens by a factor of $\sim2$ for more porous low-latitude ice, while it extends to higher latitudes. In the absence of increased porosity or reduced albedo, other mechanisms are required to reproduce the observed crystallinity patterns. These include localized thermal anomaly (which was not observed; e.g., \citealt{thelen2024subsurface}) or rapid resurfacing, operating on timescales of approximately tens of days \citep{cartwright2025jwst}.

In contrast, mid-latitude terrains outside the chaos regions are better described by a comparatively homogeneous crystalline subsurface capped by an irradiation-amorphized skin, in agreement with our thermophysical modeling (Figure~\ref{fig: amorphization_crystallization_rates}) and previous inferences on Europa's crystallinity structure \citep{cartwright2025jwst, mergny2025blinking}.

\subsection{Carbon dioxide on Europa's leading hemisphere}

The CO$_2$ absorption band is more pronounced in a lens-like pattern centered on Tara, Powys, and Annwn Regiones. The 4.25~µm band (left absorption peak of band~g), associated with CO$_2$ in porous or weakly bound matrices, strengthens overall and especially relative to the 4.27~µm band, which traces more stabilized CO$_2$ embedded in water ice (right absorption peak of band~g) \citep{he201812co2, schiltz2024characterization}. The narrow 2.7~µm $\nu_1+\nu_3$ combination absorption feature (band~c) follows a similar spatial pattern, peaking in southern Tara Regio and to a lesser extent in Powys Regio, consistent with CO$_2$ complexed within H$_2$O-dominated and mixed matrices \citep{ehrenfreund1999laboratory, white2012laboratory}. Absorption by $^{13}$CO$_2$ (band~h) is only marginally detected but increases in the same regions. 

The persistence of CO$_2$ and related species on Europa motivates mechanisms beyond simple atmospheric loss: once CO$_2$ is supplied to the atmosphere, it is removed by photoionization and magnetospheric processes on timescales of years to decades \citep{trumbo2023distribution, villanueva2023endogenous}. Laboratory studies and modeling for comets and Kuiper Belt Objects show that trapping in pore networks or amorphous domains can increase effective volatile residence times by orders of magnitude relative to compact surfaces \citep{bar1985trapping, Klinger1991, ayotte2001effect}, including Gyr timescales under cold conditions ($\sim$40~K; \citealt{birch2024retention}). In Tara Regio, where the full suite of ice-texture diagnostics is internally consistent, the observations are most compatible with a porous, structurally complex near-surface layer. Retention within such ice is a physically motivated hypothesis for sustaining localized CO$_2$ enrichment after a potential endogenous emplacement.

The lenses associated with the Powys and Annwn Regiones are located in substantially lower-albedo material than Tara Regio. Since both are found in regions adjacent to the trailing hemisphere, this is consistent with a stronger contribution from radiolytically processed non-ice components, including hydrated sulfur-bearing compounds \citep{carlson2002sulfuric, carlson2005distribution}. The muted 3.1~µm Fresnel double-lobe in those low-albedo terrains is likewise consistent with an optically altered surface veneer, even where low-latitude temperatures might otherwise favor rapid recrystallization.

However, the available ice-band suite for these observing geometries does not permit the same multi-band, self-consistent porosity conjecture as in Tara Regio. Therefore, we do not attempt to strongly link their albedo properties to the microphysics or surface conditions. Nevertheless, some ice diagnostics appear to vary spatially with these lenses (see Appendix~\ref{app_multi}).
Furthermore, the CO$_2$ band morphology offers additional constraint on the environment: in the exterior lenses, the 4.27~µm component is weakest relative to the broader 4.25~µm absorption, favoring CO$_2$ in weakly bound, mixed host environments and, potentially, structurally disordered or porous matrices \citep{he201812co2, schiltz2024characterization}. 

Interpreting localized CO$_2$ enrichment patterns across distinct chaos units as a consequence of near-surface retention properties, without necessarily sharing a single formation pathway, is a conceptually tractable hypothesis. Chaos terrains were shown to span a continuum of morphologies \citep{leonard2018analysis}, in which disrupted plates are embedded within a lumpy, often `sponge-like' matrix material that is frequently associated with dark, hydrated components \citep{collins2009chaotic}. Stratigraphic work likewise supports multiple chaos-forming episodes and local control by factors such as grain size and non-ice abundance \citep{parro2016timing}.
Together with chaos-formation models spanning brine-infiltration and refreezing processes \citep{schmidt2011active} to melt-through endmembers \citep{obrien2002melt}. These results motivate a conservative interpretation: CO$_2$ lenses need not uniquely fingerprint a single resurfacing mechanism at each site. 

Our results underscore that CO$_2$ enrichment in the chaos terrains is closely correlated with anomalous ice texture and low thermal inertia, where observed. We propose that prolonged residence of volatiles trapped within disordered, fine-grained, and potentially porous near-surface ice provides an additional mechanism for that enrichment. In this framework, the retention efficiency of a structurally complex regolith modulates volatile residence times and, therefore, contributes to the extant spatial distribution of CO$_2$, with direct implications for how Europa's oceanic materials are expressed at the surface.

Independent compositional tracers reinforce this view. Irradiated NaCl, plausibly linked to endogenous supply, is predominantly concentrated within Tara Regio, with a more limited expression in regions where additional CO$_2$ enrichment lenses are detected \citep{trumbo2019sodium}. Although these features may probe distinct effective optical depths, the lack of spatial covariance between NaCl and CO$_2$ implies that salt exposure and volatile enrichment are not governed by a single shared process spanning different chaos units.

\section{Conclusion and outlook}
This study demonstrates that spatial–spectral factorization provides a compact and physically interpretable representation of Europa's resolved reflectance. By separating intrinsic spectral structure from viewing geometry and measurement noise, the method isolates coherent modes of variability that can be compared consistently across bands and projections. The decomposition, therefore, exposes correlations that are not apparent from individual diagnostics considered in isolation.

The principal physical insight is that volatile enrichment and ice texture are spatially coupled. The data support a scenario in which near-surface microstructure modulates the observable expression of volatiles, suggesting that surface composition cannot be interpreted independently of the regolith's physical state. In this view, enrichment patterns need not uniquely trace emplacement processes; they may also reflect location-dependent retention efficiency governed by ice structure.

The present analysis does not resolve the relative contributions of emplacement and retention, but it establishes an observational approach for distinguishing them. Extending this framework to additional hemispheres, viewing geometries, and spectral windows will enable more complete surface coverage and further reduction of projection degeneracies. Coupling decomposition-based diagnostics with thermophysical modeling will allow more quantitative constraints on volatile residence times and their dependence on microphysical evolution.

More broadly, these results emphasize that interpreting the spectral signatures on Europa's surface requires considering both composition and structure. Spectral–spatial decomposition provides a systematic path toward that goal.

\begin{acknowledgments}
We are thankful to Yohai Kaspi and Barak Zackay for their support throughout this work. We are thankful to Volker Perdlewitz for assisting with the data reduction 
and to Antoine Dumont for the helpful discussions on NIRSpec properties and systematics. We also thank Oded Aharonson, Aviv Ofir, Cyril Mergny, and Guillaume Cruz-Mermy for useful discussions, and the anonymous referee for a careful and constructive review that improved the clarity of the manuscript.
S.S. acknowledges support from the Benoziyo Prize Postdoctoral Fellowship at the Weizmann Institute of Science, during which the present project was initiated. G.Y. acknowledges support from the Dean's, Faculty, and Clore Prize Postdoctoral Fellowships at the Weizmann Institute of Science.
\end{acknowledgments}

\begin{contribution}

Both authors contributed equally to this work.

\end{contribution}

%
\appendix


\section{Data processing}
\label{sec: obs and reduction}

\subsection{Data description}
We analyzed JWST NIRSpec–IFU observations of Europa's leading hemisphere obtained as part of cycles one and two programs \#1250 and \#9230, with sub-observer points at 2.63$^\circ$N, 93.61$^\circ$W; 2.63$^\circ$N, 156.87$^\circ$W; and 2.63$^\circ$N, 38.89$^\circ$W. A two-point dither pattern was used, with two 268-second exposures at each position. For calibration, we used the G0V star GSPC P330-E, a close solar analog observed under similar instrument settings on 2022 August 7 at 19:13 UT, as part of program \#1538 \cite{gordon2022james}. All data were obtained from the Mikulski Archive for Space Telescopes (MAST) at the Space Telescope Science Institute and are available at \dataset[doi: 10.17909/bxmt-4s28]{https://doi.org/10.17909/bxmt-4s28}.

We used a custom reduction pipeline designed for solar system objects \citep{king2023custom}, developed based on the official JWST Calibration Pipeline.\footnote{Version 1.17.1 with CRDS context \texttt{jwst\_1427.pmap}.} We reduced the measurements, combining the dithered exposures while avoiding despiked and desaturated products. All standard processing steps were retained, except fringe correction. Residual correlated noise and scattered light were removed using \texttt{nsclean} \citep{rauscher2024nsclean}.
The data acquired with the F100LP/G140H and F290LP/G395H filter/grating observation modes experienced detector saturation in later readout groups due to the high flux levels from Europa's disk. To preserve spectral linearity and avoid persistence artifacts, we retained only Group~1 from each integration for the final calibrated product. Group~1 corresponds to the first non-destructive read in the \texttt{MULTIACCUM} ramp, representing the earliest linear regime of the detector response before any potential saturation or non-linearity.
The spectroscopic reduction procedure included background subtraction, flat-fielding, and flux calibration using the latest reference files. Outlier rejection and resampling were performed with the default \texttt{emsm} method to align dithered exposures while preserving spectral fidelity. Wavelength calibration was performed using the pipeline solution without further refinement, and the solution was validated using known absorption features in the calibrator spectrum.

Flux calibration followed standard procedures. Uncertainties were derived from the pipeline-generated variance extensions, including contributions from detector readout and photon noise.
Finally, we identify and mitigate resampling artifacts caused by PSF subsampling in the NIRSpec IFU. We mitigate the impact of these artifacts using a low-rank filtering procedure, as described in Appendix~\ref{sec: wiggles} below.

\subsection{Rectification of the spectrum}
To prepare the data for analysis, we applied two preprocessing steps: flattening the spatial grid and rectifying the spectral dimension. 
The centroid of Europa was determined from the right ascension and declination metadata in the reduced spectral cube, which specifies the target's apparent position on the sky at the time of observation. These coordinates were cross-referenced with ephemeris data from JPL Horizons\footnote{\url{https://ssd.jpl.nasa.gov/horizons/app.html}} to verify consistency and anchor the image orientation relative to Europa’s predicted center. 
The latitude and longitude associated with each spaxel were determined by mapping detector pixels to planetocentric coordinates using the accompanying \texttt{\_nav.fits} files.

To isolate diagnostic spectral features, we applied a band-specific rectification procedure. Each spaxel's spectrum was interpolated onto a common wavelength grid and converted to reflectance using a solar analog spectrum. Uncertainties in the calibrated reflectance spectra were propagated using standard error propagation for a ratio.

For selected bands, we then rectified the continuum trend by fitting a smooth baseline function
$C(\lambda)$ to the data in wavelength intervals adjacent to the absorption feature and dividing the observed reflectance spectrum $R(\lambda)$ by it. Per bandpass, the continuum-removed spectrum is therefore given by $R(\lambda)/C(\lambda)$. This normalization suppresses large-scale continuum curvature, isolating the depth, width, and shape of the molecular absorption features.

Continuum removal was applied to the following bands using second-order polynomial fits to
adjacent continuum regions: H$_2$O at 1.65~µm (1.50--1.59~µm, 1.68--1.72~µm),
CO$_2$ at $\sim$2.7~µm (2.65--2.68~µm, 2.71--2.78~µm), CO$_2$ (4.19--4.20~µm, 4.30--4.45~µm),
and $^{13}$CO$_2$ (4.375--4.39~µm, 4.385--4.40~µm). 

\subsection{Flattened matrix form} \label{model}
\begin{figure*}
\centering
\rotatebox[origin=c]{0}{
\includegraphics[scale = 0.75,trim={5cm 0 5cm 0},clip]{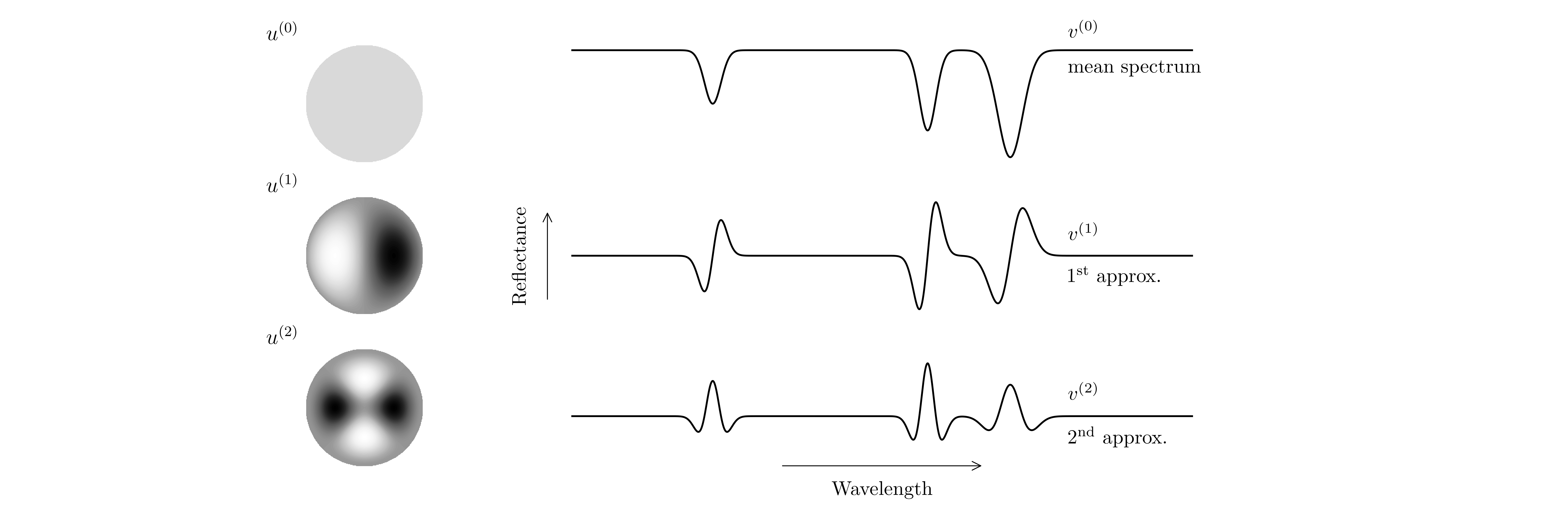}}
\caption{Qualitative illustration of the decomposition results. The first three spatial modes (columns of $U$, rearranged and approximated as spherical harmonics) are shown on the left, ordered from top to bottom by the magnitude of their associated singular values. The grayscale colormap reflects the spatial scaling of the corresponding spectral mode at each position. The principal spectra (columns of $V$) on the right represent the decomposed terms that approximate the principal spectral components. The top spectral mode shows a mock absorption spectrum, while the middle and bottom panels illustrate perturbations to this shape. For illustrative purposes, each spatial mode is approximated by a single real spherical harmonic $R_{lm}$: $R_{00}$, $R_{11}$, and $R_{22}$, from top to bottom. In practice, however, each spatial mode, $u^{(i)}$, is represented by a linear combination of multiple $R_{lm}$ terms, such that the mode number and leading spherical harmonic order are not necessarily correlated.}
     \label{fig: svd modes}
\end{figure*}
The rectified and normalized data cubes described above are the input to the decomposition analysis.
To facilitate matrix-based operations, the 2D spatial grid of size $n_{\rm col}\times n_{\rm row}$ was flattened into a one-dimensional array according to the indexing
\begin{equation}
k = i_{\mathrm{row}} \cdot n_{\mathrm{col}} + i_{\mathrm{col}},
\label{eq: flattening}
\end{equation}
where $(i_{\mathrm{row}}, i_{\mathrm{col}})$ denotes the spaxel position on the detector and $n_{\mathrm{col}}$ is the number of columns. This indexing preserves spatial locality, allowing the inverse transformation to restore the original spatial configuration when needed.

The flattened IFU data are represented as a real-valued matrix $S \in \mathbb{R}^{m \times n}$, where each row corresponds to a spaxel on the detector and each column to a wavelength bin. The spatial axis contains $m \equiv n_{\rm row} \cdot n_{\rm col}$ spaxels, with each row recording the flux across $n$ wavelength bins at a given position. This matrix is decomposed, as shown in equation~(\ref{eq: SVD}). This process yields three matrices, $V$, $U$, and $\Sigma$, which correspondingly contain the derived principal spectra, $v^{(i)}$, spatial modes, $u^{(i)}$, and their corresponding singular values, $\Sigma_{00}>\Sigma_{11}>\Sigma_{22}\dots >0$. See the definition of these quantities in the main text. 

Figure~\ref{fig: svd modes} shows an illustrative reconstruction. The right panels show mock principal spectra, while the left illustrate a spherical harmonics projection of the corresponding spatial modes. In practice, each spatial mode is typically described by a combination of a few spherical harmonics.

\subsection{Mitigating subsampling artifacts}
\label{sec: wiggles}

A known systematic effect in NIRSpec–IFU data arises from the detector undersampling of the point-spread function (PSF). This undersampling introduces quasi-periodic spectral distortions, commonly referred to as ``wiggles'' \citep{shajib2025texttt, dumont2025wiggle}. These appear as oscillatory modulations along the spectral dimension of individual spaxels and can reach amplitudes comparable to subtle astrophysical signals. If uncorrected, such artifacts can bias spectral decompositions by imprinting spurious modes that mimic genuine variability.

We mitigate these effects by filtering the principal spectra based on their spectral frequency content, rejecting components dominated by oscillatory behavior inconsistent with broadband reflectance variability. Previous work has shown that this systematic exhibits a prominent harmonic component at frequencies of order 10 cycles~µm$^{-1}$. After identifying potentially contaminated modes, we corroborate their non-astrophysical nature by noting that they are associated with small-scale spatial structure that cannot be robustly interpreted, irrespective of physical origin.
The procedure is applied directly to the reflectance spectra, without continuum rectification. The spectra are decomposed using singular value decomposition (SVD), and higher-order components identified as instrumental are removed. This is implemented by reconstructing a low-rank approximation of the spectrum, truncating modes that fail the spectral-frequency criterion.

\begin{figure*}
\centering
\rotatebox[origin=c]{0}{\includegraphics[scale = 0.52]{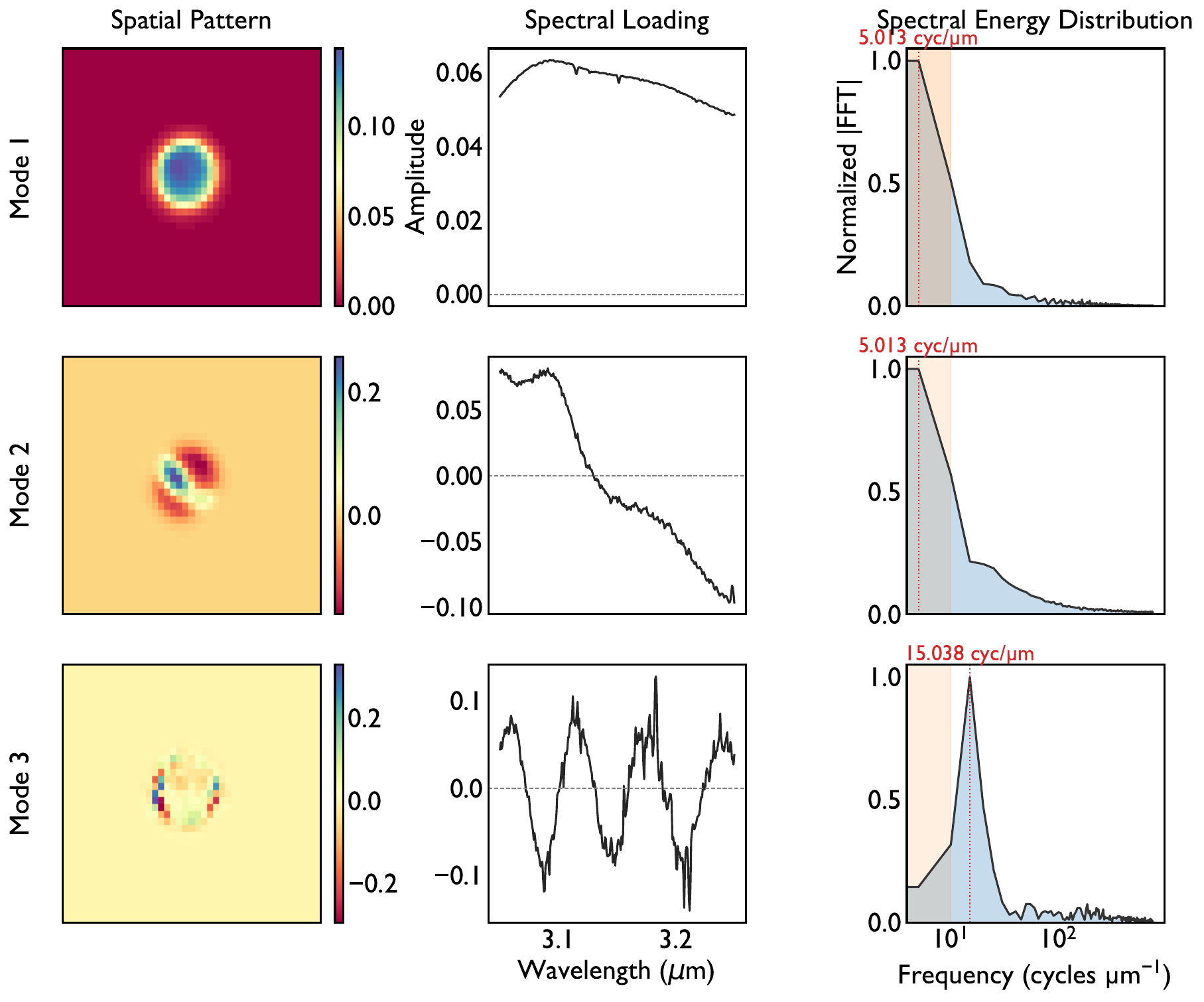}}
\caption{Diagnostic decomposition of the 3.1~µm Fresnel peak on Europa’s leading hemisphere.  Shown are the first four singular modes, with their spatial maps (left), spectral loadings (middle), and normalized FFT power spectra (right).  
The zeroth and first modes capture the broadband continuum and Fresnel feature variability, while higher–order modes display excess high–frequency power characteristic of PSF–induced wiggles.  
According to the truncation criterion, only the first two modes are retained.}
     \label{fig: svd_filter}
\end{figure*}

Figure~\ref{fig: svd_filter} illustrates this procedure for the Fresnel peak on the leading hemisphere. The first four singular vectors are shown, together with their spatial patterns, spectral loadings, and normalized FFT power spectra. The zeroth mode captures the broadband continuum and absorption variability, while the first mode isolates the dominant spectral modulation associated with the Fresnel feature and the continuum curvature. Higher-order modes exhibit shifted peak frequencies and excess high-frequency power relative to the reference window, consistent with PSF-induced wiggles. 

This criterion usually results in retaining the first two modes for all spectral bands. The spectrum is then reconstructed using only these modes, and the resulting low-rank approximation is used for subsequent analysis. The filtered spectrum corresponds to a rank-$k$ SVD reconstruction, 
\begin{equation}
\mathbf{S} \approx \sum_{i=0}^{k-1} \sigma_i\, \mathbf{u}_i \mathbf{v}_i^{\mathsf T} +{\rm systematics} + {\rm  noise},
\label{eq:svd_trunc}
\end{equation}
with $k=2$. See the example in Figure~\ref{fig: svd_filter}.
After the initial decomposition, the first term on the right-hand side is used to reconstruct the cleaned data cube. Because the low-rank reconstruction projects the data onto a restricted subspace, it absorbs any variance aligned with the retained modes, including a fraction of the stochastic white noise. The filtered spectrum, therefore, does not preserve the original noise statistics. For this reason, uncertainties are not estimated from the SVD reconstruction residuals but are instead propagated from an external shot-noise model evaluated prior to filtering (see the detailed description below).

The subsampling systematics correction is applied to the data cube before calibration with the comparison star.
An analogous correction is not required for the one-dimensional calibrator spectrum. Once the spectrum is sampled across multiple pixels within the extraction aperture, the pixel-scale phase variations of the subsampling wiggles average out \citep[e.g.,][]{dumont2025wiggle}.

\begin{figure*}
\centering
\rotatebox[origin=c]{0}{\includegraphics[scale = 0.7]{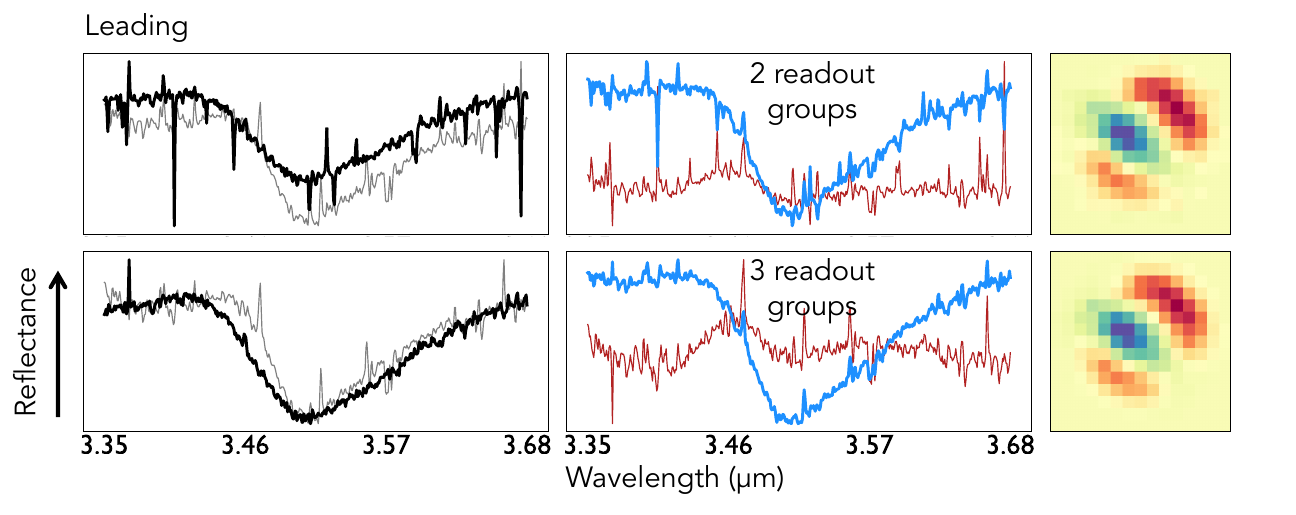}}
\caption{Normalized spectral and spatial modes of the H$_2$O$_2$ band on Europa's leading hemisphere, similar to Figure~3, analyzed over two and three readout groups. The blue and red curves illustrate $v^{(0)}\pm v^{(1)}$, respectively.}
     \label{fig: H2O2_trailing_spectrum}
\end{figure*}

An exception to this process is the feature near 3.5~µm. This signal appears broadband and spatially coherent. On the leading hemisphere, it manifests as an apparent narrowing of the H$_2$O$_2$ absorption that could have been mistaken for a false-positive variability signal in the H$_2$O$_2$ band. In Figure~\ref{fig: H2O2_trailing_spectrum}, we show the extracted feature relative to the continuum for 2- and 3-group readouts on the leading hemisphere. Under 2-group sampling, the feature is strongly diminished, consistent with flux leakage associated with the broadened diffraction pattern of bright sources \citep{dumont2025wiggle}. Therefore, pending further characterization, we treat this feature as a broadband NIRSpec artifact and refrain from analyzing the spectral band in which it is observed.

\section{Mode stability and uncertainty propagation
} \label{app: error propagation}
\subsection{Uncertainties}
Measurement noise propagates into the decomposed vectors because, in practice, the decomposition is applied not to the ideal signal matrix $S$, but to the perturbed matrix $S + E$. 
The error matrix $E$ is not directly observed and is affected by the low-rank systematic correction. Instead, each entry in $S$ has an associated one-sigma uncertainty, arranged in the matrix $\mathcal{E} \in \mathbb{R}^{m \times n}$. We assume these uncertainties are uncorrelated, as the IRS2 readout mode suppresses detector-correlated noise and the \texttt{nsclean} algorithm reduces vertical banding and background non-uniformities \citep{rauscher2024nsclean}.
The signal is shot-noise dominated, and count levels in central spaxels are sufficiently high to justify applying Gaussian statistics. Peripheral regions affected by low-count Poisson noise or residual artifacts are excluded. Correlations along the spectral axis are also considered negligible: the narrowest feature analyzed, the $^{13}$CO$_2$ line, has a full width of $\sim$0.01~µm, spanning multiple resolution elements.

\subsection{Error propagation}
The additive noise affects both the singular values and singular vectors, with the extent of this effect governed primarily by the spectral norm $\|E\|$. This norm also bounds the deviation between the perturbed and unperturbed singular values, as stated by Weyl's inequality \citep{ORourke2018}. When the entries of $E$ are modeled as independent, zero-mean, sub-Gaussian random variables, high-probability bounds on $\|E\|$ can be derived \citep{Vershynin2018}. 
Taking both steps, 
\begin{equation} 
\label{eq: wedin_bound}
|\Sigma'_{ii} - \Sigma_{ii}| \leq \|E\| \lesssim C \sigma_\star \big(\sqrt{m} + \sqrt{n}\big) \,,
\end{equation} 
where $\Sigma_{ii}$ and $\Sigma'_{ii}$ denote the $i$\textsuperscript{th} singular value matrices of $S$ and $S'$, respectively. The expression on the right-hand side depends on the matrix dimensions, $m$ and $n$; the upper bound on the standard deviation of all matrix entries, $\sigma_\star$; and on some scaling constant, $C$. In the low-rank regime relevant here, the dominant singular values should remain well above the noise level, and so should the differences between them. If this is the case, the leading singular vectors are stable: they point roughly in the same direction they would have in the ideal noise-free case, and keep the same order. 

The propagation of measurement errors into the orientation of the spatial modes (i.e., the left singular vectors) can be bounded using the Davis–Kahan–Wedin theorem \citep{ORourke2018, ORourke2024}. The angle $\Theta$ between the perturbed and unperturbed singular vectors satisfies, with high probability, 
\begin{equation} 
\label{eq: wedin_sinTheta}
\sin \Theta \leq \frac{2\|E\|}{\delta} \lesssim 
2C \frac{\sigma_\star}{\delta} \big(\sqrt{m} + \sqrt{n}\big), 
\end{equation} 
$\delta$ is the spectral gap between singular values, and $\sigma_\star$ is the standard deviation of the noise. In our case, $\Theta$ corresponds to the angle between the first-order singular vectors, $u^{(1)}$ and $u'^{(1)}$ and their corresponding spectral gap is the difference between the first and second singular values, $\delta_{12}\equiv\Sigma_{11} - \Sigma_{22}$. Since singular vectors are defined up to a sign, $\Theta$ is confined to $[0, \pi/2]$. 

Assuming that noise primarily induces a rotation of the dominant singular vectors, the Davis–Kahan–Wedin bound allows us to approximate the uncertainty in the recovered spatial modes. Since the angular deviation is bounded by $\sin \Theta$, the fractional error in the first spatial mode can be estimated as 
\begin{equation} 
\label{eq:delta_u}
\Delta u^{(1)}_k \sim \frac{1}{\sqrt{m}}\|\Delta u^{(1)}\|_2 \simeq \frac{1}{\sqrt{m}} \tan \Theta \, , 
\end{equation} 
where $m$ is the length of the spatial mode vector and $k=1,\dots,m$ is the spaxel index. This expression estimates the typical relative deviation per element due to noise-induced rotation. Because $u^{(1)}$ is a unit vector, the estimate is dimensionless and directly applicable to the least-squares inference below. It provides a conservative, interpretable upper bound on the propagation of uncertainty from the IFU measurement to the principal modes.

We performed a bootstrap simulation for each spectral band to estimate $\|E\|$. In each iteration, a noise matrix $E \in \mathbb{R}^{m \times n}$ was generated with entries $E_{ij} \sim \mathcal{N}(0, \mathcal{E}_{ij}^2)$. The spectral norm $\|E\|$ was computed and averaged across realizations to obtain a final estimate.
Table~\ref{tab:svd_noise} summarizes the spectral gaps, estimated noise norms, and corresponding upper bounds on the angular deviation of the leading singular vectors, $\sin\Theta$, using equation~(\ref{eq: wedin_sinTheta}). We adopt this upper bound as a conservative estimate of rotation in the dominant mode.

Five of the seven selected bands show small sensitivity to perturbation, with $\max(\sin \Theta)$ below ${\sim}\,0.075$. The remaining two appear to be more sensitive to noise.
The 1.65~µm band exhibits a significant rotation angle of $\max(\sin \Theta) \approx 0.42$, likely reflecting increased uncertainty due to early saturation and the use of a single readout group. For the $^{13}$CO$_2$ band, the estimated angular deviation reaches $\max(\sin \Theta) \approx 0.8$, indicating that this vector is perturbed, rendering the error estimate in equation~(\ref{eq:delta_u}) limited.

However, this could be due to a drawback in our error estimate. The $\sin \Theta$ upper limit estimates the combined impact of errors across all spatial modes, potentially affecting the estimate of a highly localized signal. The first-order spatial mode of $^{13}$CO$_2$ is localized and strongly associated with the chaos terrains, tracing the behaviour of other bands (see Figure~\ref{fig: spectrum_decomposition}). Furthermore, the detection of the zeroth-order mode is robust (as $\delta_{01}\simeq 10^3 \delta_{12}$ in this band). 
Therefore, while we consider the $^{13}$CO$_2$ first-order variability tentative, we do not rule out the possibility that it represents a genuine singular component in the spectrum. 

\begin{table*}
\centering
\caption{Maximal propagated 1$\sigma$ error, singular value gaps, spectral norm of noise (with propagated uncertainty), and angular deviation between singular vectors for each spectral band. Subscripts on $\delta_{i,j}$ indicate the pair of modes used to compute the singular value gap. $\hat{\sigma}_\star$ denotes the estimated typical propagated uncertainty, defined as the largest 1$\sigma$ propagated measurement error within each band (i.e., $\max(\mathcal{E}_{\rm band})$).}
\begin{tabular}{lccccc}
\toprule
Band & $\hat{\sigma}_\star$ [Jy] & $\delta_{01}$ [Jy] & $\delta_{12}$ [Jy] & $\|E\|$ [Jy] & $\max(\sin\Theta)$ \\
\midrule
(a) H$_2$O at 1.65~µm      & 0.19548 & $9.66044\times10^2$ & 6.74063 & $1.45048 \pm 0.07470$ & 0.4304 \\
(b) H$_2$O at 2~µm         & 0.03344 & $8.35935\times10^2$ & 34.60747 & $0.37790 \pm 0.01137$ & 0.0218 \\
(c) CO$_2$ at 2.7~µm         & 0.00573 & $6.13085\times10^1$ & 0.61466 & $0.05189 \pm 0.00237$ & 0.1688 \\
(d) H$_2$O at 3~µm              & 0.05878 & $5.03164\times10^1$ & 5.68338 & $0.11048 \pm 0.01921$ & 0.0389 \\
(e) H$_2$O Fresnel Peak     & 0.00095 & $4.68358\times10^1$ & 0.82283 & $0.01639 \pm 0.00024$ & 0.0398 \\
(g) CO$_2$ at 4.25--4.27~µm        & 0.00197 & $3.10155\times10^1$ & 0.55015 & $0.01821 \pm 0.00024$ & 0.0662 \\
(h) $^{13}$CO$_2$           & 0.00160 & $1.73957\times10^1$ & 0.04040 & $0.01613 \pm 0.00039$ & 0.7985 \\
($\zeta$) Continuum I       & 0.00094 & $2.08398\times10^1$ & 0.37134 & $0.01069 \pm 0.00024$ & 0.0576 \\
($\eta$) Continuum II       & 0.00780 & $7.87924\times10^1$ & 1.17592 & $0.02898 \pm 0.00070$ & 0.0493 \\
CO$_2$ at 4.25--4.27~µm (3D) & 0.00085 & $5.08512\times10^{1}$ & 0.68617 & $0.02050 \pm 0.00023$ & 0.05975 \\
H$_2$O Fresnel Peak (3D) & 0.00034 & $7.60287\times10^{1}$ & 1.46205 & $0.01759 \pm 0.00024$ & 0.02406 \\

\bottomrule
\end{tabular}
\label{tab:svd_noise}
\end{table*}

\section{Sparse Spherical Harmonic Approximation} \label{app: model_Ylm}
\subsection{Single projected hemisphere}
To model the spatial structure of the dataset, we construct a linear model using a superposition of \textit{real} spherical harmonics. Any spatially varying function defined on the sphere can be expanded in terms of these basis functions, given up to a normalization factor by
\begin{equation}
R_{\ell m}(\theta, \phi) \propto
\begin{cases} 
\cos(m \phi) P_{\ell}^{m}(\cos\theta) & \text{if } m \geq 0, \,\,\text{and}\\[8pt]
\sin(|m| \phi) P_{\ell}^{|m|}(\cos\theta) & \text{if } m < 0.
\end{cases}
\end{equation}
The normalization factor is omitted, as the basis functions are explicitly normalized before fitting. Spherical harmonics are defined on the surface of a three-dimensional unit sphere. To apply them to detector-plane observations, we use an orthographic projection. We define a local Cartesian coordinate system $(y, z)$ centered on Europa’s centroid in the image plane. The viewing direction is aligned with the $x$-axis, the north pole points along the $ z$-axis, and $R$ denotes the projected radius of Europa. Under these assumptions, the transformation between Cartesian and spherical coordinates is
\begin{equation}
\begin{aligned}
    r &= \sqrt{y^2 + z^2},         &\quad x &= \sqrt{R^2 - r^2}, \\
    \theta &= \arccos\left({x}/{R}\right), &\quad \phi &= \arctan2(z / y).
\end{aligned}
\end{equation}
All positions and radii are measured in physical spaxel units on the detector. In practice, positions on the detector plane are discretized and translated into a running index for the first spatial mode. 

Using equation~(\ref{eq: flattening}), each spaxel on the observed hemisphere is assigned a positional index $k$. For spaxels within the projected disk ($r \leq R$), this index maps to angular coordinates via the chain
\begin{equation}
    k \rightarrow (y_k,z_k) \rightarrow (\theta_k, \phi_k, r_k).
\end{equation}
Accordingly, for spaxels within Europa's projected hemisphere, the first spatial mode is modeled as
\begin{equation}
u^{(1)}_k = \sum_{\ell=0}^{\ell_{\rm max}} \sum_{m=-\ell}^{\ell} a_{\ell m} R_{\ell m}(\theta_k, \phi_k)  + \text{noise}.
\end{equation}
We assume the signal vanishes outside the projected disk and that the measurements contain only additive noise. 
Europa spans approximately 13 spaxels in projection, corresponding to a disk area of $\sim$133 spaxels. Accounting for PSF smoothing and limb foreshortening, we estimate $60$–$80$ effective spatial degrees of freedom. Consequently, we adopt $\ell_{\text{max}} = 7$ as the cutoff of our expansion. This truncation yields 64 coefficients, matching the resolution and remaining within the Nyquist limit.

\begin{figure*}[t!]
\centering
\rotatebox[origin=c]{0}{\includegraphics[scale = 0.75]{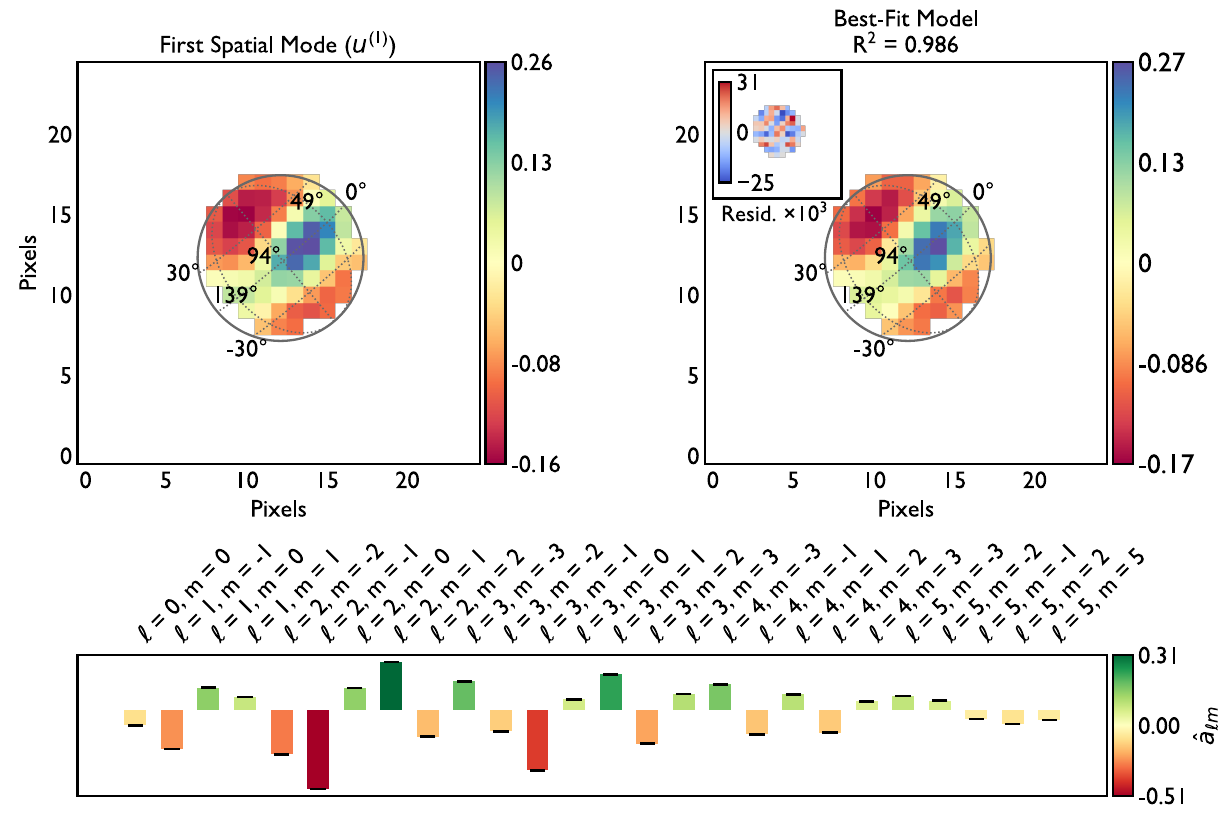}}
\caption{Illustration of the fitting process of the first spatial mode, $u^{(1)}$, of the H$_2$O Fresnel Peak. Left: Observed spatial map corresponding to $u^{(1)}$. Right: Best-fit spherical harmonic model reconstruction using coefficients up to $\ell_{\rm{max}} = 5$, with inset showing residuals. The approximate centroid and projected circumference are delineated with a dashed black line. Bottom: best-fit spherical harmonic mode coefficients ($\hat{a}_{\ell m}$), color-coded by amplitude. Error bars denote propagated uncertainties.}
     \label{fig: fit_example}
\end{figure*}

The coefficients $a_{\ell m}$ are determined by observed data, and the residual noise term reflects measurement uncertainty, estimated using equation~(\ref{eq:delta_u}).
The model is linear in the parameters; therefore, it can be solved using the normal equations. We define the design matrix $X$ as
\begin{equation} \label{eq:design_matrix}
    X_{kj} = 
    \begin{cases}
        R_{\ell m}(\theta_k, \phi_k) & \text{if } r_k\leq R, \\
        0 & \text{otherwise},
    \end{cases}
\end{equation}
where $k$ is the positional index and $j$ enumerates over the fitted $R_{\ell m}$ harmonics. This construction ensures that the fitted functions are non-zero only in the physically meaningful region, corresponding to Europa's projection on the detector.
However, because the projection eliminates line-of-sight information, different spherical harmonic modes can map onto indistinguishable patterns in the image plane. As a result, the projected basis functions are no longer orthogonal. We perform Gram-Schmidt orthonormalization on the columns of $X$.
The Gram-Schmidt process is then applied to $X$ to yield a design matrix of orthonormal basis functions,
\begin{equation}
    X_{o} = \mathscr{G}(X),
\end{equation}
where $\mathscr{G}$ denotes the Gram-Schmidt process, and $X_o$ is the design matrix containing the orthonormalized basis functions used in the decomposition model. 
Solving the normal equations yields the fitted coefficients. Since the entries of $u^{(1)}$ are assumed uncorrelated with identical variance, and due to the orthonormality of the columns of $X_o$, the solution of the normal equations attains a simple form,
\begin{equation}
\hat{\mathbf{a}} = \left( X_o^\top  X_o\right)^{-1} X_o^\top  {\mathbf{u}}^{(1)}  \pm \frac{\tan \Theta}{\sqrt{m}} \mathbf{1}_m,
\end{equation}
where $\mathbf{1}_m$ is a vector of ones of length $m$, and the scaling of the uncertainty is due to equation~(\ref{eq:delta_u}).
Figure~\ref{fig: fit_example} illustrates the fitting procedure of the first spatial mode associated with the H$_2$O Fresnel Peak. The non-uniform variance distribution across spherical harmonic modes highlights the compactness of the reconstructed signal and its dominant angular scales.

\subsection{Multiple viewing angles}
When multiple viewing angles are available, we perform a simultaneous fit to infer a single set of spherical harmonic coefficients that describes all projected views. In this appendix, we consider only the first-order spatial mode. For simplicity, we drop the mode index and denote the first-order spatial mode by $u^{(v)}$, where the superscript labels the viewing angle; the formulation extends straightforwardly to higher-order modes.

Each observation is associated with a design matrix $X^{(v)}$, constructed according to equation~(\ref{eq:design_matrix}) by evaluating the spherical-harmonic basis at the geometry-dependent angles $(\theta_{k,v}, \phi_{k,v})$. The spherical coordinates are defined in Europa's planetocentric coordinate system, with a common zero point adopted for all viewing angles.
We fit a single set of spherical-harmonic coefficients to all viewing angles. Constraints from the individual viewing angles are therefore combined into a single linear system by stacking the corresponding spatial modes and concatenating their design matrices,
\begin{equation}
\mathbf{u} \equiv 
\begin{bmatrix}
u^{(1)}\\
\vdots\\
u^{(n_v)}
\end{bmatrix},
\qquad
X \equiv 
\begin{bmatrix}
X^{(1)}\\
\vdots\\
X^{(n_v)}
\end{bmatrix}.
\end{equation}
In contrast to the single–viewing-angle case, the Gram-Schmidt procedure is not applied to the design matrices. Preserving a common spherical-harmonic basis is required to infer a single coefficient vector. Per-view orthonormalization would break this consistency.

We infer the spherical-harmonic coefficient vector $a$, truncated at degree $\ell_{\max}$. The resulting linear inverse problem is intrinsically ill-conditioned: each observation samples only a fraction of the surface, leaving large regions, including the poles and trailing hemisphere, unconstrained. We therefore solve for $a$ using Tikhonov-regularized weighted least squares \citep{golub1999tikhonov} by minimizing the objective function
\begin{equation}
\mathcal{L}(\mathbf{a})
=
\mathbf{r}^\top W \,\mathbf{r}
+\lambda\,\mathbf{a}^{\top}\Lambda \,\mathbf{a},
\label{eq:multi_geom_tikhonov_w}
\end{equation}
where $\mathbf{r}\equiv \mathbf{u} - X\mathbf{a}$ is the residuals vector, $W$ is the inverse-variance weight matrix, and $\lambda$, which controls the regularization strength, is set to 0.01 in our analysis.
We adopt a Laplacian regularizer in the spherical-harmonic basis,
\begin{equation}\label{eq: regularization term}
\mathbf{a}^\top \Lambda\,\mathbf{a} \equiv \sum_{lm} C \,\ell(\ell+1)\,a_{\ell m}^2,
\end{equation}
suppressing poorly constrained small-scale structure. $C$ is a normalization constant, discussed below. The resulting closed-form solution is
\begin{equation}
\hat{\mathbf{a}}=\Bigl(X^{\top}WX+\lambda\,\Lambda\Bigr)^{-1}X^{\top}W  \mathbf{u}.
\end{equation}

We set the inverse-variance weights using a uniform per-spaxel uncertainty estimated from the noise-induced rotation of the first-order spatial mode (Appendix~\ref{app: error propagation}). As shown there, the rotation of an orthonormal eigenvector can be bounded by an angle $\Theta$, implying a typical spaxel-wise perturbation of order $\tan\Theta/\sqrt{n_{\rm{eff}}}$. See equations~(\ref{eq: wedin_sinTheta})~and~(\ref{eq:delta_u}).

In the multi-geometry case, we apply this estimate to the stacked system. We compute $\Theta$ from the concatenated band matrix used for the joint decomposition (equation~\ref{eq: SVD multi hemisphere}) and replace $m$ by $n_{\mathrm{eff}}$, the effective number of spaxels contributing to $\mathbf{u}$. We then define a diagonal inverse-variance weight matrix with elements
\begin{equation}
W_{ii} \equiv \frac{n_{\mathrm{eff}}}{\tan^2\Theta}.
\end{equation}
This choice assumes uniform uncertainty across spaxels and aims to provide a simple, interpretable weighting scheme. The scaling constant from equation~(\ref{eq: regularization term}) is set to the same value, $C = W_{ii}$, to ensure that the regularization term scales with the likelihood. In practice, the noise may be spatially non-uniform. In such cases, resampling-based approaches, such as bootstrap methods, may be more appropriate. However, for the purposes of this work, where projection onto a spherical-harmonic basis provides a sparse, smooth representation rather than quantitative surface comparisons, these estimates are sufficient. 

\begin{figure*}[t!]
\centering
\rotatebox[origin=c]{0}{\includegraphics[scale = 0.55]{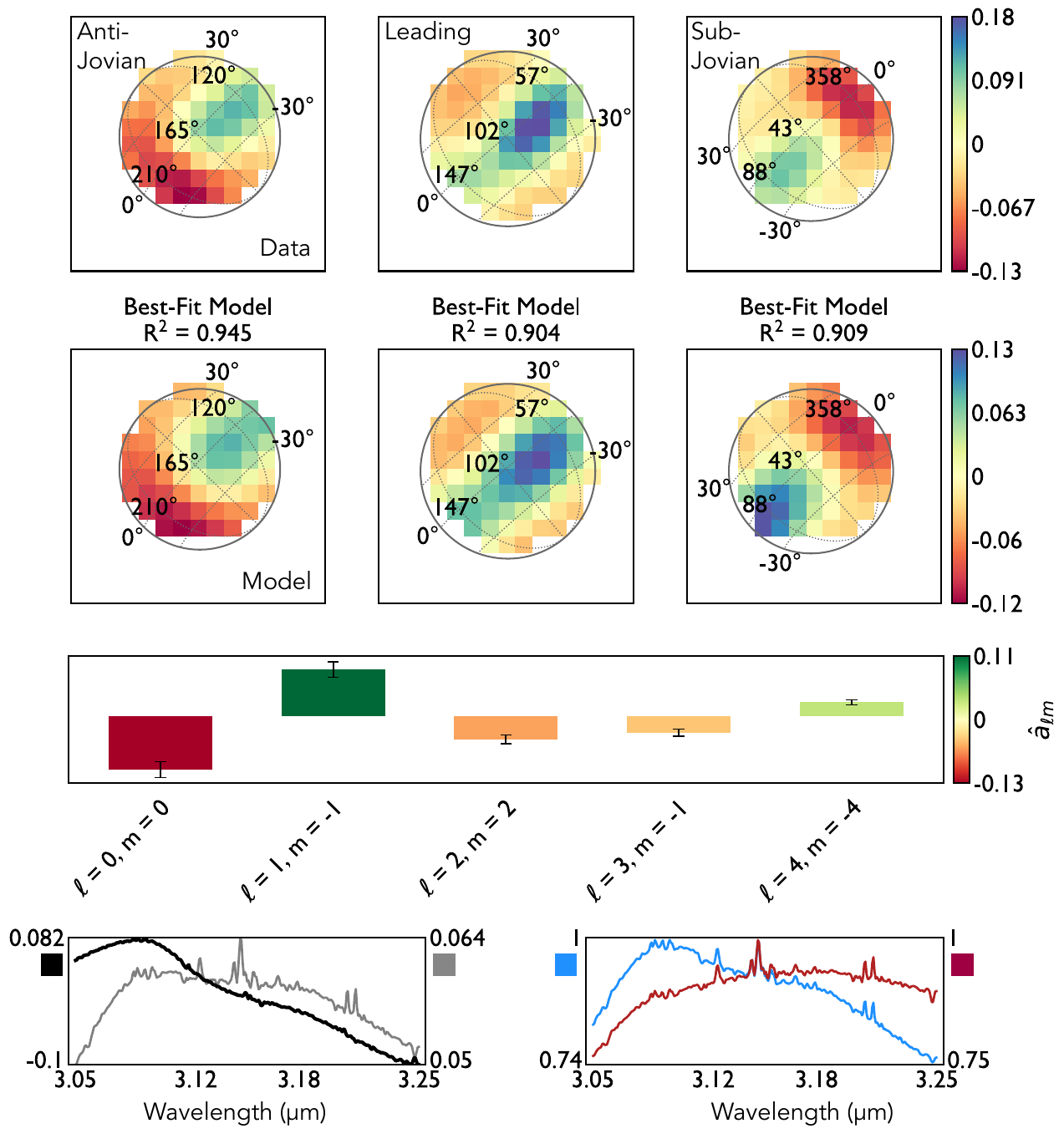}}
\caption{Illustration of the simultaneous multi‑geometry fit for the first spatial mode, $u^{(1)}$, of the 3.1~µm H$_2$O Fresnel peak ($\ell_{\text{max}} = 6$). Top row: observed $u^{(1)}$ maps for each viewing geometry (columns). Middle row: best‑fit spherical harmonic reconstructions from a joint fit across all geometries, with per‑view $R^2$; the projected limb and latitude/longitude grid are overplotted. Bottom: dominant joint best‑fit spherical harmonic coefficients ($|\hat{a}_{\ell m}| \geq 0.1\cdot\rm{max}(|\hat{a}_{\ell m}|)$), color‑coded by amplitude with propagated uncertainties.}
     \label{fig: multigeometry_fit_example}
\end{figure*}

We select the spherical-harmonic truncation $\ell_{\max}$ empirically to balance expressiveness against noise amplification. Specifically, we increase $\ell_{\max}$ until each viewing geometry, analyzed independently, achieves a coefficient of determination of at least $R^2 \ge 0.9$ for the fitted spatial mode, and adopt the smallest $\ell_{\max}$ that satisfies this criterion for all geometries (Figure~\ref{fig: multigeometry_fit_example}). This fit described here essentially represents a change of basis. The spherical-harmonic expansion provides a compact parameterization for comparing projections in a common coordinate system, while residual limitations are confined to polar regions where incomplete sampling reduces sensitivity.

\section{Multi-Geometry Fits in the 2--3 Micron Window} \label{app_multi}

\begin{figure*}[t!]
\centering
\rotatebox[origin=c]{0}{\includegraphics[scale = 0.26]{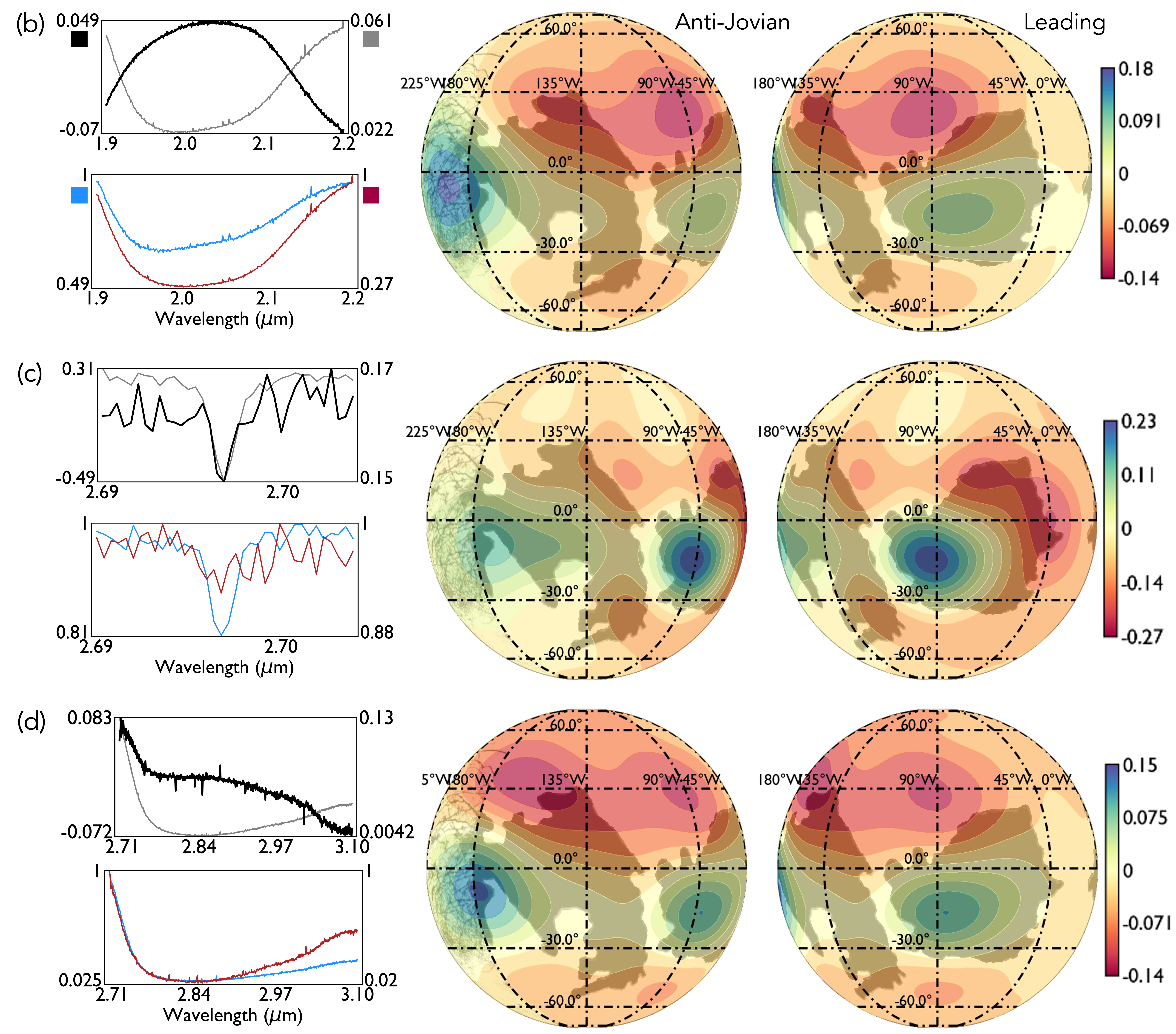}}
  \caption{Qualitative projections of the best-fit three-dimensional spherical-harmonic model to the jointly decomposed first spatial mode, $u^{(1)}$, derived from two viewing geometries of Europa's leading hemisphere. (b) The 2.7~µm $\nu_1+\nu_3$ CO$_2$ absorption band ($\ell_{\rm{max}}=20$). (c) The $\sim$3~µm H$_2$O fundamental ($\ell_{\rm{max}}=11$). (d) The $\sim$2~µm H$_2$O absorption band ($\ell_{\rm{max}}=11$). Shaded outlines mark Tara and Powys Regiones. The left panels show jointly decomposed spectral modes ($v^{(0)}$ and $v^{(1)}$ in gray and black, respectively) and spectra from the spaxels where the joint spatial mode peaks (blue) and reaches its minimum (red).
}
  \label{fig: 2-3_micron_multproj}
\end{figure*}

We analyze JWST NIRSpec--IFU observations of Europa acquired with the G235H/F170LP configuration in Cycles~1 and~2 (programs \#1250 and \#4023), providing access to the $\sim$2--3~µm range in both the leading and anti-Jovian viewing geometries. All reduction and calibration steps follow Appendix~\ref{sec: obs and reduction}; the only additional handling specific to this configuration is retaining only the first readout group of each MULTIACCUM ramp to mitigate saturation and persistence in Europa’s high-flux regime. Figure~\ref{fig: 2-3_micron_multproj} shows the multi-geometry analysis of these bands, analogous to Figure~\ref{fig: fresnel_co2_mult_proj}, and carried out as described in Section~\ref{sec: multi_geometry_fit}. Table~\ref{tab:svd_noise_app} summarizes the singular-value gaps, estimated noise norms, and the corresponding upper bounds on angular deviation for the jointly decomposed leading singular vectors, analogous to Table~\ref{tab:svd_noise}.

\begin{table*}
\centering
\caption{Maximal propagated 1$\sigma$ error, singular value gaps, spectral norm of noise (with propagated uncertainty), and angular deviation between singular vectors for each spectral band, similarly to Table~\ref{tab:svd_noise}.}
\begin{tabular}{lccccc}
\toprule
Band & $\hat{\sigma}_\star$ [Jy] & $\delta_{01}$ [Jy] & $\delta_{12}$ [Jy] & $\|E\|$ [Jy] & $\max(\sin\Theta)$ \\
\midrule
H$_2$O at 2~µm (3D)        & 0.03345 & $1.01683\times10^3$ & 49.36371 & $0.39403 \pm 0.01276$ & 0.0160 \\
CO$_2$ at 2.7~µm (3D)         & 0.00574 & $5.20197\times10^1$ & 0.24391 & $0.05297 \pm 0.00137$ & 0.4343 \\
H$_2$O at 3~µm (3D)        & 0.03815 & $6.90661\times10^1$ & 3.09489  & $0.08622 \pm 0.01014$ & 0.0557 \\
\bottomrule
\end{tabular}
\label{tab:svd_noise_app}
\end{table*}

\section{Ice Evolution Model} \label{app_ice}

Two counteracting processes determine whether water ice is in a crystalline or amorphous phase: The first is thermal annealing, where temperature-dependent kinetics arrange the ice into a crystalline lattice \citep{jenniskens1996crystallization, mitchell2017porosity}. The second is radiation-induced amorphization, which drives the destruction of the crystalline lattice via deposition of radiative energy \citep{strazzulla1992ion, fama2010radiation}.
Since both insolation and radiative energy deposition rates are depth- and location-dependent, near-surface ice phase on Europa has also been shown to strongly depend on its geographic location \citep{berdis2020europa, yoffe2025fluorescent}. 

To interpret the observed spatial patterns in ice crystallinity (see Figure~\ref{fig: projections}a, Figure~\ref{fig: projections}$\zeta$, and Figure~\ref{fig: projections}$\eta$), we modeled the competing effects of radiolytic amorphization and thermal recrystallization under realistic surface conditions. This allowed us to estimate whether the observed spectral signatures can arise from steady-state physical processes or require localized anomalies such as recent resurfacing or enhanced porosity.
We assessed the water-ice phase within the uppermost 10 microns of Europa's near-surface at several latitudes between 0$^\circ$ and 30$^\circ$. For this purpose, we estimated the independent timescales for amorphization and crystallization at different locations and depths. We considered two kinds of ice, which, in this model, differed only in density: porous ice ($\rho_{\rm{ice}} = 275.1$ kg m$^{-3}$) and compact ice ($\rho_{\rm{ice}} = 500$ kg m$^{-3}$) \citep{mitchell2017porosity, mergny2025blinking, thelen2024subsurface}. The model was applied similarly to both kinds. 
Two inputs were required to evolve crystallization and amorphization over time: (i) diurnally cycling temperature profiles and (ii) radiative energy deposition rates.

To estimate the diurnal temperature cycle at different latitudes and depths, we solved a one-dimensional vertical surface heat equation, following \cite{ashkenazy2019surface}. For the radiative energy deposition rates, we adopted latitude- and depth-resolved profiles from simulations presented in \cite{yoffe2025fluorescent}.
We computed the decoupled timescales required to reach crystalline and amorphous ice fractions for each latitude and depth element at each time step.
Crystallization was evaluated via the Avrami formalism \citep{jenniskens1996crystallization, mitchell2017porosity}, and amorphization was modeled as a time-dependent process driven by the cumulative radiative energy deposited into the ice. Specifically, we used a simplified exponential model in which the amorphous fraction increases with total dose according to an empirically derived efficiency constant for light ions---the dominant energy depositing species across Europa's leading hemisphere \citep{fama2010radiation, yoffe2025fluorescent}. Both the crystallization and amorphization timescales are defined as the duration required to achieve an $e$-fold (approximately 63\%) fraction. 

Figure~\ref{fig: amorphization_crystallization_rates} compares the amorphization and crystallization timescales across latitude and depth for porous and highly-porous ice. 
Porous ice absorbs a higher radiation dose per unit depth than more compact ice, as the dose scales inversely with density (but the stopping power of ions does not). However, its lower thermal inertia makes it more susceptible to higher diurnal temperatures. This accelerates thermal recrystallization compared to irradiation-driven amorphization, as recrystallization rates rise exponentially with temperature, following an Arrhenius relationship \citep{mitchell2017porosity}.
Consequently, for both kinds of ice, crystallization outpaces amorphization at all depths up to latitude 30$^\circ$, but the highly-porous ice recrystallizes at faster rates, primarily in low-latitude regions. Beyond latitude $\sim$40$^\circ$, amorphization progressively overtakes in the uppermost near-surface layers.
A detailed account of all the constituents of the ice evolution model is provided below.

\subsection{Diurnal Temperature Cycle} \label{app_ice_thermal}
To estimate the diurnal thermal structure of Europa’s uppermost ice, we adopt the standard one-dimensional surface heat diffusion model of \citet{ashkenazy2019surface}. The model solves for the time-dependent temperature profile under periodic solar forcing, radiative losses, and a constant upward internal heat flux. We integrate the model over the upper 10 cm of ice with sufficient vertical resolution to resolve the diurnal skin depth and evolve it for 200 days to reach a periodic steady state. Surface albedo values are taken from the high-resolution Bond albedo map of \citet{mergny2025bond}, averaged over two representative polygons of Tara Regio and the mid-latitudes. All thermal parameters are summarized in Table~\ref{tab:thermal_parameters}.

\begin{table}[h]
\centering
\caption{Thermal model parameters adopted from *\cite{ashkenazy2019surface}, **\cite{spencer1999temperatures}, ***computed using the Bond albedo map from \cite{mergny2025bond}, as described above, and $\dagger$ice density values chosen for observed mean and elevated porosities (at leading hemisphere chaos terrains) of $\approx$0.5 and $\approx$0.7, respectively \cite{thelen2024subsurface, mergny2025blinking}.}
\label{tab:thermal_parameters}
\tiny
\begin{tabular}{lll}
\hline
\textbf{Symbol} & \textbf{Description} & \textbf{Value} \\
\hline
$\rho_{\rm{ice}}$$\dagger$ & Ice density (porous, highly-porous) & 500, 275.1~kg\,m$^{-3}$ \\
$c_{p,I}$* & Heat capacity of ice & 2000~J\,kg$^{-1}$\,K$^{-1}$ \\
$\epsilon$** & Emissivity of ice & 0.9 \\
$\alpha_p$*** & Bond albedo (Tara Regio, mid-lat) & 0.64, 0.65 \\
$\kappa_s$* & Thermal diffusivity of surface ice & $7.7 \times 10^{-10}$~m$^2$\,s$^{-1}$ \\
$\sigma$ & Stefan–Boltzmann constant & $5.67 \times 10^{-8}$~W\,m$^{-2}$\,K$^{-4}$ \\
$p$* & Eclipse fraction & 0.033 \\
$W_J$* & Longwave flux from Jupiter & 0.176~W\,m$^{-2}$ \\
$Q$* & Internal heat flux & 0.05~W\,m$^{-2}$ \\
$T_{\rm{init}}$* & Initial temperature profile & 60~K \\
\hline
\end{tabular}
\end{table}

\subsection{Magnetospheric Particle Energy Deposition}
\label{app_ice_Edep}

To model the radiation environment at Europa’s surface, we adopt the depth-resolved energy deposition profiles presented in \citet{yoffe2025fluorescent}. These simulations use \texttt{G4beamline} \citep{roberts2007g4beamline} to track charged particle transport through near-surface ice, accounting for electrons and magnetospheric ions (p, O$^{2+}$, and S$^{3+}$). Particle power spectra are constrained by \textit{Voyager} and \textit{Galileo} measurements and modulated by magnetospheric drift patterns appropriate for Europa’s leading hemisphere \citep{paranicas2001electron, nordheim2018preservation, nordheim2022magnetospheric}.

The resulting energy deposition rates are provided as a function of depth (down to one meter) and surface location. For full details of the particle transport simulations and assumptions, we refer the reader to \citet{yoffe2025fluorescent}, Appendix~A.\footnote{The depth- and location-resolved dose rate maps are available online in Zenodo \citep{https://doi.org/10.5281/zenodo.18944511}.}

\subsection{Ice Crystallization and Amorphization}
\label{app_ice_phase}

The near-surface ice structure on Europa is governed by the competition between thermally driven crystallization and radiation-induced amorphization. Both processes were evaluated using laboratory-based kinetic formalisms under diurnally varying temperature and radiation environments.

Crystallization of amorphous ice was modeled using Avrami kinetics following \citet{jenniskens1996crystallization, mitchell2017porosity}. The crystalline fraction evolves as
\begin{equation}
f_c(t)=1-\exp\!\left[-\int_{0}^{t} n\,K\!\big(T(\tau)\big)\,\tau^{\,n-1}\,d\tau\right],
\end{equation}
where $n=2$ is appropriate for heterogeneous nucleation in porous systems. The crystallization rate constant follows an Arrhenius relation \citep{vyazovkin2022jeziorny},
\begin{equation}
K(T)=K_0^n\,\exp\!\left(-\frac{n\Delta H}{RT}\right),
\end{equation}
with an Avrami prefactor $K_0 = 3.9\times 10^{19}$ s$^{-1}$ and activation enthalpy $\Delta H = 60$ kJ mol$^{-1}$ for porous ice \citep{mitchell2017porosity}.

Radiation-driven amorphization was modeled using a dose-dependent exponential formalism based on laboratory irradiation experiments \citep{fama2010radiation, baragiola2013radiation}. The amorphous fraction evolves as
\begin{equation}
f_a(t)=1-\exp\!\left[-\int_{0}^{t} k\!\big(T(\tau)\big)\,Dd\tau)\right],
\end{equation}
where $D(t)$ is the cumulative deposited energy dose (in eV per molecule), and $k(T)$ is the temperature-dependent amorphization efficiency derived from experimental measurements of ion- and electron-induced amorphization \citep{strazzulla1992ion, loeffler2020possible}.

For each latitude and depth, diurnally cycling temperature profiles and local energy deposition rates (Section~\ref{app_ice_Edep}) were used to evaluate both processes simultaneously. Crystallization timescales were defined as the time required to reach $f_c=1-e^{-1}\simeq0.63$, while amorphization proceeded continuously through cumulative dose deposition. The relative efficiency of the two processes determines the steady-state ice structure at each location.

\bibliography{sample}{}
\bibliographystyle{aasjournalv7}

\end{document}